\documentclass[conference]{IEEEtran}

\usepackage{subfigure}
\usepackage{balance}
\usepackage{graphicx}
\usepackage{amsmath}
\usepackage{amssymb}
\usepackage{algorithm}
\usepackage[noend]{algorithmic}
\usepackage{multirow}
\usepackage{url}
\usepackage{xcolor}
\usepackage{soul}
\usepackage{xspace}

\usepackage{cleveref}
\newcommand{\stitle}[1]{\vspace{1ex}\noindent{\bf #1}}

\newcommand{\ie}{{\em i.e.,}\xspace}

\title{
Multi-Attribute Selectivity Estimation Using Deep Learning
}

\author
 {
 \IEEEauthorblockN{
     Shohedul Hasan$^{\ddag}$,
     Saravanan Thirumuruganathan$^{\ddag\dag}$,
	 Jees Augustine$^{\ddag}$,
     Nick Koudas$^{\ddag\ddag}$, 
     Gautam Das$^{\ddag}$
     }
 \IEEEauthorblockA{
     $^{\ddag}$University of Texas at Arlington;
     $^{\ddag\dag}$QCRI, HBKU;
     $^{\ddag \ddag}$University of Toronto
 }
 {
     $^{\ddag}$\{shohedul.hasan@mavs,~jees.augustine@mavs,~gdas@cse\}.uta.edu,
     $^{\ddag\dag}$sthirumuruganathan@hbku.edu.qa,
     $^{\ddag\ddag}$koudas@cs.toronto.edu
 }
}

\begin{document}

\maketitle

\begin{abstract}
Selectivity estimation -- the problem of estimating the result size of queries -- is a fundamental problem in databases.
Accurate estimation of query selectivity involving multiple correlated attributes is especially challenging.
Poor cardinality estimates could result in the selection of bad plans by the query optimizer.
We investigate the feasibility of using deep learning based approaches for both point and range queries
and propose two complementary approaches.
Our first approach considers selectivity as an unsupervised deep density estimation problem.
We successfully introduce techniques from neural density estimation for this purpose.
The key idea is to decompose the joint distribution into a set of tractable conditional probability distributions such that they satisfy the autoregressive property.
Our second approach formulates selectivity estimation as a supervised deep learning problem that predicts the selectivity of a given query.
We also introduce and address a number of practical challenges arising when adapting deep learning for relational data.
These include query/data featurization, incorporating query workload information in a deep learning framework and the dynamic scenario where both data and workload queries could be updated.
Our extensive experiments \emph{with a special emphasis on queries with a large number of predicates and/or small result sizes}
demonstrates that our proposed techniques provide fast and accurate selective estimates with minimal space overhead.
\end{abstract}

\section{Introduction}
\label{sec:intro}

Selectivity estimation -- the problem of estimating the result size of queries with multiple predicates --
is a fundamental yet challenging problem in databases.
It has diverse applications in query optimization, query profiling, database tuning, approximate query processing etc.
Poor cardinality estimates could result in the selection of bad plans by the query optimizer~\cite{leis2015good}.
Due to its importance, this problem has attracted intense interest from the database community.

\stitle{Current Approaches and their Limitations.}
Accurate estimation of query selectivity involving multiple (correlated) attributes is especially challenging.
Exactly representing the joint distribution is often infeasible when many attributes are involved or each attribute could take large number of values.
Broadly speaking, major database systems tackle this problem by approximating this joint distribution via synopses or sampling.
Synopsis techniques such as histograms approximate the joint frequency distribution in a bounded space by making assumptions
such as uniformity and attribute value independence~\cite{poosala1996improved,leis2015good}.
These assumptions are often violated in real-world datasets resulting in large errors in selectivity estimation~\cite{leis2015good}.
Building multidimensional histograms could partially ameliorate this issue but often has substantial space requirements.
Sampling based approaches could handle attribute dependencies and correlations more effectively.
However, it is not a panacea -- for queries with low selectivity, the optimizer could be made to rely on magic constants~\cite{leis2015good}, resulting in poor estimates.

\subsection{Outline of Technical Results}
\label{subsec:technicalResults}
In this paper, we investigate the suitability of Deep Learning for selectivity estimation.
Building a DL model that is lightweight, fast to train and estimate,
and optionally allow injection of domain knowledge such as query  workload is non trivial.
We propose two complementary approaches that operate in two phases.
In the offline phase, we train an appropriate DL model from the data.
During the online phase, the model accepts a query and outputs its selectivity.

\stitle{Selectivity Estimation as Unsupervised Learning.}
Our first approach models selectivity estimation as a density estimation problem
where one seeks to estimate the joint probability distribution from a finite number of samples.
Intuitively, the traditional sampling and synopses approaches can be considered as approximate non-parameteric density estimators.
However, instead of directly estimating the joint probability,
we seek to decompose it into a series of simpler and tractable conditional probability distributions.
Specifically, we consider a specific decomposition with \emph{autoregressive} property (formally defined in Section~\ref{sec:unsupervised}).
We then build a single DL model to simultaneously learn the parameters for each of the conditional distributions.

\stitle{Selectivity Estimation as Supervised Learning.}
We investigate if,
given a training set of queries along with their true selectivity,
is it possible to build a DL model that accurately predicts the selectivity of unknown queries involving multiple correlated attributes?
Our proposed approach can be utilized to quickly train a model \emph{without} having seen the data!
The training set of queries with their true selectivities could be obtained from the query logs.
Our DL models are lightweight and can provide selectivity estimates for datasets in few milliseconds.
Our model outperforms other supervised estimation techniques such as Multiple Linear Regression and Support Vector Regression
that have been applied for the related problem of query performance prediction~\cite{akdere2012learning}.
The key benefit factor is the ability of DL models to handle complex non linear relationships between queries involving correlated attributes and their selectivity.

\stitle{Summary of Experiments.}
We conducted an extensive set of experiments over two real-world datasets -- Census and IMDB -- that exhibit complex correlation and conditional independence between attributes and have been extensively used in prior work~\cite{leis2015good}.
We specifically focus on queries that have multiple attributes and/or small selectivity.
We evaluated our supervised and unsupervised DL models on a query workload of 10K queries.
Our supervised model was trained on a training data of 5K queries.
Our results demonstrate that DL based approaches provide substantial improvement - for a fixed space budget - over prior approaches
for {\em multi-attribute selectivity estimation} which has been historically a highly challenging scenario in database selectivity estimation.

\stitle{Summary of Contributions.}
\begin{itemize}
	\itemsep-1mm
	\item \stitle{Deep Learning for Selectivity Estimation.}
		We introduce an alternate view of database selectivity estimation namely as an neural density estimation problem and report highly promising results.
		Our algorithms could handle queries with both point and range predicates.
	\item \stitle{Making the approach suitable for Databases.}
		We describe adaptations making these models suitable for various data types, large number of attributes and associated domain cardinalities,
		availability of query workload and incremental queries and data.
	\item \stitle{Experimental Validation.}
		We conduct extensive experiments over real-world datasets establishing that our approach provides accurate selectivity estimates for challenging queries, including the challenging cases of queries involving large number of attributes.
\end{itemize}

\stitle{Paper Outline.}
Section~\ref{sec:preliminaries} introduces relevant notations. 
Section~\ref{sec:unsupervised} formulates selectivity estimation as an unsupervised neural density estimation problem
and proposes an algorithm based on autoregressive models.
In Section~\ref{sec:supervised}, we introduce the problem of selectivity estimation and propose a supervised Deep Learning based model for it.
Section~\ref{sec:experiments} describes our extensive experiments on real-world datasets,
related work in Section~\ref{sec:relWork} and finally conclude in Section~\ref{sec:futureWork}.

\section{Preliminaries}
\label{sec:preliminaries}

\subsection{Notations}
Let $R$ be a relation with $n$ tuples and $m$ attributes $A=\{A_1, A_2, \ldots, A_m\}$.
The domain of the attribute $A_i$ is given by $Dom(A_i)$.
We denote the value of attribute $A_i$ of an arbitrary tuple as $t[A_i]$.
We consider conjunctive queries with both point and range predicates.
Point queries are of the form $A_i = a_i \text{ AND } A_j = a_j \text{ AND } \ldots$
for attributes $\{A_i, A_j\} \subseteq A$ where $a_i \in Dom(A_i)$ and $a_j \in Dom(A_j)$.
Range queries are of the form $lb_i \leq A_i \leq ub_i \text{ AND } lb_j \leq A_j \leq ub_j \text{ AND } \ldots$
Let $q$ denote such a conjunctive query while $Sel(q)$ represents the result size.
We use the normalized selectivity between $[0,1]$ by dividing the result size by $n$, number of tuples.

\stitle{Performance Measures.}
Given a query $q$, let the estimate provided by selectivity estimation algorithm be $\widehat{Sel(Q)}$.
We use \emph{q-error} for measuring the quality of estimates.
Intuitively, q-error describes the factor by which the estimate differs from true selectivity.
This metric is widely used for evaluating selectivity estimation approaches~\cite{leis2015good,leis2017cardinality,kipf2018learned,muller2018improved}
and is relevant for applications such as query optimization where the relative ordering is more important~\cite{leis2015good}.
We do not consider the use of relative error due to its asymmetric penalization of estimation error~\cite{muller2018improved}
that results in models that systematically under-estimate selectivity.
\begin{equation}
	\text{q-error} = \max\left( \frac{Sel(Q)}{\widehat{Sel(Q)}}, \frac{\widehat{Sel(Q)}}{Sel(Q)} \right)
\end{equation}


\subsection{Selectivity Estimation as Distribution Estimation}
\label{subsec:selectivityAsDistrEst}

Given a set of attributes $A'=\{A_i, A_j, \ldots,\}$, the normalized selectivity distribution defines a valid (joint) probability distribution.
The selectivity of a query $q$ with $\{A_i=a_i, A_j=a_j, \ldots\}$ can be identified by locating the appropriate entry in the joint distribution table.
Unfortunately, the number of entries in this table increases exponentially in the number of attributes and their domain cardinality.

Distribution estimation is the problem of learning the joint distribution from a set of finite samples.
Often, distribution estimators seek to approximate the distribution by making simplifying assumptions.
There is a clear trade-off between accuracy and space.
Storing the entire distribution table produces accurate estimates but requires exponential space.
On the other hand, heuristics such as attribute value independence (AVI) assume
that the distributions of individual attributes $A_i$ are independent of each other.
In this case one needs to only store the individual attribute distributions and
compute the joint probability as
$$p(A_i=a_i, A_j=a_j, \ldots) = \prod_{A_k \in A'} p(A_k = a_k)$$

Of course, this approach fails for most real-world datasets that exhibit correlated attributes.
Most popular selectivity estimators such as multidimensional histograms, wavelets, kernel density estimations and samples
can be construed as simplified non-parametric density estimators on their own.

\subsection{Desiderata for DL Estimator}
Given that selectivity estimation is just one component in the larger query optimization framework,
we would like to design a model that aids in the identification of good query plans.
Ideally, the estimator should be able to avoid the unrealistic assumptions of uniformity and attribute value independence (afflicting most synopses based approaches)
and ameliorate issues caused by low selectivity queries (afflicting sampling based approaches).
We would like to decouple training-accuracy tradeoff.
For example, increasing the sample sizes improves the accuracy - at the cost of increasing the estimation time.
If necessary, the estimator could have a large training time to increase accuracy but should have near constant estimation time.
We would also like to decouple the space-accuracy tradeoff.
Multi-dimensional histograms can provide reasonable estimates in almost constant time - but require very large space (that grows exponentially to the number of attributes) for accurate results.
In other words, we would like to achieve high accuracy through a lightweight model.
The desired model must be fast to train and given the latency requirements of query optimizer, generate estimates in milliseconds.
It must also be able to appropriately model the complex relationship between queries and their selectivities.
Finally, it must be able to leverage additional information such as query workload and domain knowledge.

\section{Selectivity Estimation as Neural Density Estimation}
\label{sec:unsupervised}

We introduce an alternate view of selectivity estimation namely as an neural density estimation problem.
This new perspective allows us to leverage the powerful tools from deep learning to get accurate selectivity estimation
while also raising a number of non-trivial challenges in wrangling these techniques for a relational database setting.

\stitle{Prior Approaches and Their Limitations.}
Past approaches to selectivity estimation include formulations as a density estimation problem.
Sampling based approaches~\cite{lipton1990practical} approximate the density of the dataset $R$ using a sample $S$.
For an uniform random sample, the normalized selectivity of query $q$ is estimated as $Sel_D(q)=Sel_S(q)$.
This approach is meaningful if the sample is large and representative enough which is not often the case.
A more promising avenue of research seeks to approximate the density through simpler distributions.
Recall from Section~\ref{subsec:selectivityAsDistrEst} that the two extremes involve
storing the entire joint distribution or approximate it by assuming attribute value independence requiring
$$\prod_{i=1}^{m} |Dom(A_i)| \text{ and } \sum_{i=1}^{m} |Dom(A_i)|$$
storage. While the former has perfect accuracy, the latter could provide inaccurate estimates for queries involving correlated attributes.
One approach investigated by the database community uses Bayesian networks (BN) that
approximates the joint distribution through a set of conditional probability distributions ~\cite{getoor2001selectivity,tzoumas2011lightweight}.
This approach suffers from two drawbacks.
First, learning the optimal structure of BN based on conditional independence is prohibitively expensive.
Second, the conditional probability tables themselves could impose large storage overhead
if the attributes have a large domain cardinality and/or are conditionally dependent on other attributes with large domain cardinality.

We address this using two key ideas:
(a) we avoid the expensive conditional independence decomposition using a simpler autoregressive decomposition;
(b) instead of storing the conditional probability tables, we \emph{learn} them.
Neural networks are universal function approximators~\cite{Goodfellow-et-al-2016}
and we leverage their powerful learning capacity to model the conditional distributions in a concise and accurate way.

\subsection{Density Estimation via Autoregressive Decomposition}
The fundamental challenge is to construct density estimators
that are expressive enough to model complex distributions while still being tractable and efficient to train.
In this paper, we focus on autoregressive models \cite{Goodfellow-et-al-2016} that satisfy these properties.
Given a \emph{specific ordering of attributes}, autoregressive models decompose the joint distribution into multiple conditional distributions
that are combined via the chain rule from probability.
Specifically,
\begin{equation}
\begin{split}
	& p(A_1=a_1, A_2, \ldots, A_m=a_m) \\
	&=\prod_{i=1}^{m} p(A_i = a_i | A_1 = a_1, A_2=a_2, \ldots, A_{i-1}=a_{i-1})
\end{split}
\end{equation}

Autoregressive models decompose the joint distribution into $m$ conditional distributions $P(A_i | A_1, \ldots, A_{i-1})$.
Each of these conditional distributions is then learned using an appropriate DL architecture.
One can control the expressiveness of this approach by controlling the DL model used to learn each of these conditional distributions.
For the attribute ordering $A_1, A_2, \ldots, A_m$,
the DL model for estimating $A_{i}$ only accepts inputs from $A_1, A_2, \ldots, A_{i-1}$.
The DL model first learns the distribution $p(A_1)$, followed by conditional distributions
such as $p(A_2|A_1)$, $p(A_3|A_1, A_2)$ and so on.
This process of sequentially \emph{regressing} each attribute through its predecessors is known as autoregression \cite{Goodfellow-et-al-2016}.

Given such a setting, we need to address two questions.
First, which DL architecture should be used to learn the autoregressive conditional distributions?
Second, how can we identify an effective ordering of attributes needed for autoregressive decomposition?
Different decompositions could have divergent performances and it is important to choose an appropriate ordering efficiently.

\subsection{Autoregressive Density Estimators}

\stitle{Encoding Tuples.}
The first step in modelling is to encode the tuples such that they can be used efficiently for density estimation by a DL model.
A naive approach would be to use one-hot encoding of the tuples.
While effective, it is possible to design a denser encoding.
Without loss of generality, let the domain of attribute $A_j$ be $[0, 1, \ldots, |Dom(A_j)|-1]$.
As an example, let $Dom(A_j) = \{v_{j1}, v_{j2}, v_{j3}, v_{j4}\} = [0, 1, 2, 3]$.
One-hot encoding represents them as $4$ dimensional vectors $0001$, $0010$, $0100$ and $1000$.
We could also use a \emph{binary encoding} that represents them as a $\lceil \log_2{|Dom(A_j)|} \rceil$ dimensional vector.
Continuing the example, we represent $Dom(A_j)$ as $00, 01, 10, 11$ respectively.
This approach is then repeated for each attribute individually and
the representation for the tuple is simply the concatenation of the binary encoding of each attribute.
This approach requires less storage - $\sum_{i=1}^{m} \lceil \log_2{|Dom(A_i)|} \rceil$ dimensions for $m$ attributes
instead of $\sum_{i=1}^{m} |Dom(A_i)|$ required by one-hot encoding.
As we shall demonstrate experimentally, binary encoding is faster to train due to a lower number of parameters to learn and yet
generates better estimates than one-hot encoding.

\stitle{Loss Function.}
Using binary encoding, we represent a tuple $t$ as a vector of binary observations $x$ of dimension $D$.
Each of these observations could be considered as a binary random variable $x_i$.
We can specify an autoregressive distribution over the binary encoding of the tuple as
\begin{equation}
	p(x) = \prod_{i=1}^{D} p(x_i | x_1, \ldots, x_{i-1})
\end{equation}
A naive approach would be to store conditional probability tables for various values of $x_1, \ldots, x_{i-1}$.
However, this approach would impose a large storage overhead.
An elegant approach is to treat the conditional distribution as a random variable that -
given values for $x_1, \ldots, x_{i-1}$ - takes a value of $1$ with probability $\widehat{x_i}$ and $0$ otherwise.
In other words, let $p(x_i = 1|x_1, \ldots, x_{i-1})=\widehat{x_i}$ and $p(x_i = 0|x_1, \ldots, x_{i-1}) = 1 - \widehat{x_i}$.
If we train a model to learn the value of $\widehat{x_i}$ accurately, then we can forego the need to store the conditional probability table.
In order to train such a ML model, we need to specify the loss function.
For a given tuple $x$, the negative log likelihood for estimating the probabilities is
\begin{equation}
	\label{eq:crossEntropyLossTuple}
	\begin{split}
		\ell(x) = -\log p(x) &= \sum_{i=1}^{D} - \log p(x_i | x_1, \ldots, x_{i-1}) \\
					& 	\qquad - (1 - x_i) p(x_i=0 | x_1, \ldots, x_{i-1}) \\
	\end{split}
\end{equation}

The negative log likelihood for the relation is specified as
\begin{equation}
	\label{eq:crossEntropyLossDataset}
	\begin{split}
		-\log p(R) &= \sum_{x \in R} \sum_{i=1}^{D} -x_i \log \widehat{x_i} - (1 - x_i) \log (1 - \widehat{x_i}) \\
					&= \sum_{x \in R} \ell(x)
	\end{split}
\end{equation}

The function $\ell(x)$ corresponds to cross entropy loss~\cite{Goodfellow-et-al-2016} that measures
how the learned probability $\widehat{x_i}$ diverges from the actual value for $x_i$.

\stitle{Autoregressive Density Estimators using MADE.}
We can now utilize any DL architecture such as a fully connected one to learn the conditional distributions
by minimizing the cross entropy loss defined over all the tuples as in Equation~\ref{eq:crossEntropyLossDataset}.
However, since the underlying loss is defined over an autoregressive distribution,
it is often more efficient and effective to use one of the neural autoregressive density estimators specifically designed for this purpose \cite{DBLP:conf/icml/GregorDMBW14,DBLP:journals/corr/UriaML13,germain2015made}.
While our approach is agnostic to the specific estimator used, we advocate for the masked autoencoder architecture from~\cite{germain2015made}.
MADE modifies the autoencoders~\cite{Goodfellow-et-al-2016}\cite{DBLP:conf/icml/VincentLBM08}\cite{DBLP:conf/icml/RifaiVMGB11} for efficiently estimating autoregressive distributions.
As we shall describe in Section~\ref{subsec:attributeOrdering}, its flexible architecture allows us to effectively adapt it to relational domains.

\subsection{Answering Range Queries}
\label{subsec:unsupervisedRangeQueries}

Once the autoregressive density estimator has been trained it could be used to answer \emph{point} queries.
Given a query $q: A_i = a_i \text{ AND }$, we can encode this query and feed it to the autoregressive density estimator model which will output the normalized selectivity.
While these models cannot directly answer range queries,
it is possible to use their ability to answer point queries in a sophisticated way
to obtain accurate range selectivity estimates.

Specifically, let us consider the question of answering range queries of the form:
$q: A_1 \in R_1 \text{ AND} A_2 \in R_2 \text{ AND } \ldots$
where $R_i$ is a range of the form $lb_i \leq A_i \leq ub_i$.
Point queries of the form $A_i = a_i$ could be made into a range query $a_i \leq A_i \leq a_i$.
Finally, if an attribute $A_i$ is unspecified then it could be modeled as
$\min(Dom(A_i)) \leq A_i \leq \max(Dom(A_i))$.

For the rest of subsection, we consider a query $Q$ of the form $A_1 \in R_1 \text{ AND } \ldots \text{ AND } A_k \in R_k$.

\stitle{Exhaustive Enumeration.}
If the range query is relative simple -- involving small number of attributes and/or small ranges --
a simplistic solution is to enumerate all possible combinations of the ranges and
invoke the point query estimator.
Specifically,
\begin{equation}
	sel(Q) = \sum_{x_1 \in R_1} \ldots \sum_{x_k \in R_k} p(x_1, \ldots, x_k)
\end{equation}

\stitle{Uniform Sampling.}
This approach is not feasible if many attributes with possibly large ranges are involved in the query.
For example, a query $A_1 \in [1, 100] \text{ AND } A_2 \in [1, 100]$ would need to enumerate over 10,000 point queries.
One possible solution is to generate random queries by uniformly sampling from from the ranges $R_1, \ldots, R_k$ resulting in specific values $a_1, \ldots a_k$.
Then one can query the point query estimator for $q_i = p(x_1 = a_1^i, \ldots, x_k=a_k^i)$.
This is then repeated for $|S|$ times to generate the selectivity estimate as
\begin{equation}
	sel(Q) = \frac{|R_1| \times \ldots |R_k|}{|S|} \sum_{i=1}^{|S|} sel(q_i)
\end{equation}

\stitle{Adaptive Importance Sampling.}
While intuitive, uniform sampling provides bad selectivity estimates
when the number (and range) of predicates increases due to curse of dimensionality~\cite{murphy2012machine}.
The key insight to improve the naive uniform sampling is to make it \emph{adaptive} and \emph{weighted}.
In other words, each sample could have a different weight and
the probability with which a new point is selected could vary based on previously obtained samples.
However, naively implementing this idea results in biased and incorrect results.

We adapt an algorithm that was originally designed for Monte-Carlo multi-dimensional integration
for the range selectivity estimation problem.
Intuitively, we wish to select samples $S$ in proportion to the contribution they make to $sel(q)$.
However, this leads to a chicken-and-egg problem as we use sampling to estimate $sel(q)$.
The solution is to proceed in stages and use the information collected from samples of previous stages to improve the next stage.

Let $f(\cdot)$ be the probability density function based on query $q$
such that if we sample points proportional to $f(\cdot)$, we might get accurate estimates.
Of course, this information is not always available.
Suppose that we have access to another simpler probability density function  $g(\cdot)$
that is an approximation of $f(\cdot)$ and is also easier to sample from.
Obviously, sampling from $g(\cdot)$ would provide much better estimates than uniform sampling.
Given sample queries $q_1, \ldots, q_k$ generated using $g(\cdot)$,
we can derive the estimate as
\begin{equation}
	sel(Q) = \frac{|R_1| \times \ldots |R_k|}{|S|} \sum_{i=1}^{|S|} \frac{sel(q_i)}{g(q_i)}
\end{equation}

Intuitively, we generated random queries based on $g(\cdot)$ and then appropriately corrected the bias to
get an unbiased estimate.
Now the remaining question is to obtain an efficient instantiation of $g(\cdot)$.
We propose a simple approach inspired by Attribute Value Independence (AVI) assumption
where
\begin{equation}
	\label{eq:vegasEstimate}
	\begin{split}
		Sel(A_1 = a_1 \text{ AND } \ldots \text{ AND } A_k = a_k) = \\
			Sel(A_1=a_1) \times \ldots \times Sel(A_k=a_k)
	\end{split}
\end{equation}

It is known that AVI assumption often provides an underestimate for correlated attributes~\cite{poosala1997selectivity}.
We leverage this fact to decompose $g(A_1, A_2, \ldots, A_k)$
 as $k$ component functions $g_1(A_1)$, $g_2(A_2) \ldots$.
One can then approximate the density of each of these attributes individually through
existing synopses approaches such as histograms.

Our proposed approach operates in stages.
We generate an initial batch of random queries through uniform sampling from the ranges.
Using these random queries, we bootstrap the histograms for individual attributes.
In the future stages, we generate samples in a non-uniform way using the sampling distribution imposed by the attribute wise histograms.
For example, a query $q_i = x_1 = a_1^i, \ldots, x_k=a_k^i$
will be picked proportional to the probability $g_1(a_1^i) \times \ldots \times g_k(a_k^i)$.
So if some value $A_i=a_i$ occurs much more frequently then it will be reflected in the histogram of $A_i$
and thereby will occur more frequently in randomly generated queries.
Once all the sample queries are created, we then use Equation~\ref{eq:vegasEstimate} to generate unbiased estimates for range selectivity.

\subsection{Attribute Ordering for Autoregression}
\label{subsec:attributeOrdering}
In practice, the best attribute ordering for autoregressive decomposition is not given to us and must be chosen appropriately for accurate selectivity estimation.
Each of the permutations of the attributes forms a valid attribute ordering and could be used to estimate the joint distribution.
\begin{equation}
	\label{eq:diffOrderings}
	\begin{split}
		p(x) 	&= p(x_1) \cdot p(x_2|x_1) \cdot p(x_3|x_1, x_2) \\
				&= p(x_2) \cdot p(x_3|x_2) \cdot p(x_1|x_2, x_3) \\
				&= p(x_3) \cdot p(x_2|x_3) \cdot p(x_1|x_2,x_3) \\
				&= \ldots
	\end{split}
\end{equation}

\stitle{Random Attribute Ordering.}
Prior approaches such as Bayesian Networks deploy an expensive approach to identify a good ordering.
We do away with this expensive step by choosing several random orderings of attributes.
As we shall show experimentally, this approach works exceedingly well in practice.
This is due to two facts:
(a) the vast majority of the $d!$ possible permutations is amenable to tractable and accurate learning; and
(b) the powerful learning capacity of neural networks (and masked encoders) can readily learn even a challenging decomposition by increasing the depth of the MADE model.
MADE architecture allows this to be easily and efficiently conducted by
randomly permuting both the input tuple that is binary encoded and the internal mask vectors in each layer.

\stitle{Ensembles of Attribute Orderings.}
While a random ordering often provides good results, it is desirable to guard against the worst case scenario of a bad permutation.
We observe from Equation~\ref{eq:diffOrderings} that numerous attribute orderings could be used for estimating the joint distribution.
We build on this insight by choosing $\kappa$ random attribute orderings.
Of course, different orderings result in different models with their corresponding estimate and associated accuracy.
MADE could be used to learn the conditional distribution for each of these orderings
and utilize them to estimate the value of $p(x)$ by averaging the individual estimates.
An attribute ordering can be represented as $m^{0} = [m^{0}(1), \ldots m^{0}(D)]$. In this, $m^{0}(d)$ represents the position of the $d$-th dimension of input ${\bf x}$ in the product of conditionals. Thus multiple random orderings can be obtained by permuting $[1, \ldots, D]$.

During training, before each minibatch~\cite{Goodfellow-et-al-2016} update of the model,
we apply $\kappa$ random permutations in parallel on the input vectors and mask matrices.
Each of these permutations corresponds to a different ordering.
The models are learned independently and the joint probability is computed for each ordering and averaged to produce the final estimate.
This ensemble approach minimizes the likelihood of a bad estimates due to an unlucky attribute ordering.

\stitle{Injecting Domain Knowledge.}
If a domain expert possesses apriori knowledge that attributes $A_i$ and $A_j$ are order sensitive,
then we only chose permutations where the desired order is observed.
As a concrete example, assume one knows (say via data profiling), that a functional dependency \emph{EmpID} $\rightarrow$ \emph{Department} exists on the schema.
Then, we would prefer permutations where the \emph{Department} occurs after \emph{EmpID}.
This is due to the fact that the conditional distribution $p(\text{Department}|\text{EmpID})$ is simpler and thereby easier to learn than the other way around.

\subsection{Incorporating Query Workload}
\label{subsec:MADEQueryWorkload}

The autoregressive approach outlined above does not require a training dataset such as a query workload.
However, it is possible to improve performance by leveraging query workload if available.
Suppose that we are given a query workload $Q = \{q_1, \ldots, q_l\}$.
We associate a weight $w(t)$ for each tuple $t \in R$ that corresponds to the number of queries that match $t$.
So $w(t)$ can vary between $0$ and $l$.
Next, we assign higher penalties for poor estimates for tuples in the result set of multiple queries.
The intuition is that a poor estimate for tuple $t$ was caused by
sub-optimal learning of parameter weights of the conditional distributions corresponding to the attribute values of $t$.
As an example, consider a tuple $t = [0, 1]$ with two binary attributes $A_1$ and $A_2$.
Suppose that we use a single attribute ordering $A_2, A_1$.
If the selectivity of $t$ was incorrectly estimated, then the entries corresponding to $p(A_2=1)$ and $p(A_1=0|A_2=1)$ must be improved.
If $t$ is in the result set of by many queries, then we prioritize learning the aforementioned parameter values through larger penalty.
This could be achieved using the weighted cross-entropy loss function defined as,
\begin{equation}
	\label{eq:weightedCrossEntropyLossDataset}
	\begin{split}
		-\log p(R) 	&= \sum_{t \in R} \mathbf{w(t)} \cdot \ell(t)
	\end{split}
\end{equation}

\subsection{Incremental Data and Query Workload}
\label{subsec:incrementalMADE}
We next consider the scenario where data is provided incrementally as well as new queries involving the new data become available.

\stitle{Incremental Learning.}
The naive solution of retraining the entire model from scratch becomes progressively expensive as more and more batches of incremental data are added to $R$.
We propose an incremental learning approach that \emph{extends} the existing pre-trained model by training it further only on the new data by initializing the model with the weights learned from the previous training, instead of performing the standard random initialization.
We then continue training the model on new data.
This two-step process preserves the knowledge gained from the past, absorbing knowledge from new data and is also more efficient.
We use a smaller value for learning rate~\cite{Goodfellow-et-al-2016} and epochs than for the complete retraining so that the model is \emph{fine-tuned}.

While incremental learning is conceptually simple, it must be done carefully.
A naive training could cause \emph{catastrophic forgetting} where the model ``forgets'' the old data and focuses exclusively on the new data.
This is undesirable and must be avoided~\cite{goodfellow2013empirical}.
We propose the use of Dropout~\cite{srivastava2014dropout} and related techniques~\cite{li2018learning} to \emph{learn without forgetting}.
In our paper, we utilize a dropout value of $p=0.1$ when training over the new batch of data.
An additional complication arises from the fact that we already uses masks for maintaining the autoregressive property.
Hence, we apply the dropout only on the neurons for which the masks are non-zero.

\stitle{Incremental Workload.}
An autoregressive approach does not directly utilize query workload and hence could not use the information available from an
incremental query workload.
It is possible to reuse the techniques from Section~\ref{subsec:MADEQueryWorkload} for this scenario.
For each tuple, we update the number of queries it satisfies and retrain the model based on the new weights.

\section{Selectivity Estimation as Supervised Learning}
\label{sec:supervised}


Our objective is to build a model that accepts an arbitrary query as input and outputs its selectivity.
This falls under the umbrella of supervised learning methodologies using regression.
Each query is represented as a set of features and the model learns appropriate weights for these features utilizing them to estimate the selectivity.
The weights are learned by training the model on a dataset of past queries (such as from query log or workload) and their true selectivities.
Approaches such as linear regression, support vector regression etc
that have been utilized for query performance prediction~\cite{akdere2012learning} are not suitable for building selectivity estimators.
The impediment is the complex relationship between queries and their selectivities where simplifying assumptions such as attribute value independence do not hold.
We leverage the powerful learning capacity of neural networks - with appropriate architecture and loss functions - to model this relationship.


\subsection{Query Featurization}
\label{subsec:queryFeaturization}
The first step is to encode the queries and their selectivities in an appropriate form suitable for learning.

\stitle{Training Set.}
We are given a query training dataset $Q = \{(q_1, s_1), \ldots, \}$.
Each query $q \in Q$ can be represented as an ordered list of $m$ attribute pairs of $(A_i, v_i)$
where $v_i \in Dom(A_i) \cup \{*\}$ (where * is used when $A_i$ is unspecified).
$s_i$ denotes the normalized selectivity of $q_i$ (\ie $\frac{Sel(q_i)}{n}$) where $n$ is the number of tuples.

\stitle{Example.}
Let $Q = \{( \{A_1=0, A_2=1\}, 0.3), (\{A_1=1, A_2=*\}, 0.2), (\{A_1=*,A_2=*\}, 1.0) \}$.
\ie Query $q_2$ with $A_1=1 \text{ AND } A_2=*$ has a selectivity of $0.2$.

\stitle{Encoding Queries.}
An intuitive representation for categorical attributes is one-hot encoding.
It represents attribute $A_i$ as $|Dom(A_i)|+1$ dimensional vector that has 0 for all positions except the one corresponding to the value $A_i$ takes.
Given $m$ attributes, the representation of the query is simply the concatenation of one-hot encoding of each of the attributes.
The numeric attributes can be handled by treating them as categorical attributes by automatic discretization~\cite{dougherty1995supervised}.
Alternatively, they can be specified as a normalized value $\in [0, 1]$ by min-max scaling.
Note that this scheme can be easily extended to operators other than $=$.
The only modification required is to represent the triplet $(A_i, operator_i, v_i)$ instead of just $(A_i, v_i)$.
Each operator could be represented as a fixed one-hot encoding of its own.
Given $d$ operators, each operator is represented as a $d$ dimensional vector where the entry corresponding to $operator_i$ is set to $1$.
Of course, the rest of our discussion is oblivious to other mechanisms to encode the queries.


\stitle{Encoding Selectivities.}
Each query $q \in Q$ is associated with the normalized selectivity $s_i \in [0, 1]$.
Selectivities of queries often follow a skewed distribution
where few queries have a large selectivity and the vast majority of queries have much smaller selectivities.
Building a predictive model for such skewed data is often quite challenging.
We begin by applying log transformation over the selectivity
by replacing the selectivity $s_i$ by its absolute log value as $abs(\log(s_i))$.
For example, a selectivity of $0.00001$ is specified as $5$ (using log to the base of 10 for convenience).
This has a smoothing effect on the values of a skewed distribution~\cite{changyong2014log}. 
Our second transformation is min-max scaling where we rescale the output of the log transformation back to $[0,1]$ range.
Given a set of selectivities $S = \{s_1, s_2, \ldots, \}$ and a selectivity $s_i$,
min-max scaling is computed as
\begin{equation}
	s'_i = \frac{s_i - min(S)} {max(S) - min(S)}
\end{equation}
While this transformation does not impact skew, it enables us to deploy well known activation functions such as sigmoid that are numerically stable.
Prior works such as~\cite{kipf2018learned,rangeQueriesML} have also used log transformation to improve effectiveness of regression.

\stitle{Example.}
Let the selectivities of three queries be $[0.1, 0.01, 0.002]$.
By applying the log transformation, we get $[1, 2, 3]$.
The corresponding min-max scaling gives $[0.0, 0.5, 1.0]$ where $0.5 = \frac{2 - min([1, 2, 3])} {max([1, 2, 3]) - min([1, 2, 3])}$.

\subsection{DL Model for Selectivity Estimation}

\stitle{DL Architecture.}
Our DL architecture is based on a 2-layer fully connected neural network with rectifier activation function (ReLU) specified as $f(x) = max(0,x)$
ReLU is a simple non-linear activation function with known advantages such as faster training and sparser representations.
The final layer uses a Sigmoid activation function $f(x) = \frac{1}{1 + e^{-x}}$
Sigmoid is a popular function that squashes its parameter into a $[0,1]$ range.
One can then convert this output to true selectivity by applying inverse of min-max and log scaling.
We used the Adam optimizer \cite{Kingma2015AdamAM} for training the model.

\stitle{Loss Function.}
%
%
Recall from Section~\ref{sec:preliminaries} that the q-error metric is widely used to evaluate the selectivity estimator.
Hence, it is desirable to train the DL model to directly minimize the mean Q-error of the training dataset.
\begin{equation}
	\text{q-error}(Q) = \frac{1}{|Q|} \sum_{i=1}^{|Q|} \max \left( \frac{s_i}{\widehat{s_i}}, \frac{\widehat{s_i}}{s_i} \right)
\end{equation}

\stitle{Selectivity Estimation via Inference}
Once the model is trained, it can be used for estimating the selectivity.
Given a new query, we extract its features through one-hot encoding and feed it to the model.
We apply the inverse transformation of min-max and log scaling on the output so that it represents the actual selectivity.
The model is lightweight and the inference process often takes few milli-seconds when run on a GPU and/or CPU.
Note that the time taken for training and estimation are decoupled.
While the training time is proportional to the size of the training data, the inference is fixed for a given model.

\subsection{Generating Training Data}
\label{subsec:queryWorkload}
We next describe how a training dataset could be constructed when query workload is not available.
If query workload is available, we describe a novel augmentation strategy such that the DL model
can generate accurate estimates for unknown queries that are similar to the query workload.

\stitle{No Query Workload Available.}
\label{subsubsec:queryWorkloadNotAvailable}
Naive sampling from the space of all queries results in a highly non-uniform training dataset and a sub-optimal selectivity estimator.
Thus one must obtain a training set of queries that are diverse both in the number of predicates and their selectivities.
Let the query budget be $B$ - \ie we wish to construct a dataset with $B$ queries and their selectivities.
We begin by enumerating all queries with $1$ predicates that are the atomic units from which multi-predicate queries could be estimated.
We then generate multi-predicate queries where the predicates are chosen at random while the values are chosen based on their frequency.
In order to generate a random query $q_i$, we first choose the number of predicates $k \in \{2, \ldots, m\}$ uniformly at random.
Then we choose $k$ attributes uniformly at random from the set of attributes $A = \{A_1, \ldots, A_m\}$.
Let the selected attributes be $\{A_{i1}, \ldots, A_{ik}\}$.
These two steps ensure that we have a diverse set of multi predicate queries both in terms of the number of predicates and the chosen predicates.
Next, we choose a tuple $t$ uniformly at random from the relation $R$.
We create a random query $q_i$ as the conjunction of predicates $A_{ij}  = t[A_{ij}]$.
This process ensures that the random query is selected proportional to the selectivity of query $q_i$.


\stitle{Query Workload Available.}
\label{subsubsec:queryWorkloadAvailable}
If a query workload $Q$ is available, one could directly utilize it to train the DL model.
However, one can do much better by augmenting it, obtaining a more informative training set of queries.
The key idea is to select queries from the distribution induced by the query workload
such that the model generalizes to unknown queries from the same distribution.
We need to address two issues.
First, how can one generate random queries to augment the query workload?
Second, how do we tune the model such that it provides accurate results for the workload?
The solution involves importance sampling and weighted training respectively.

We begin by assigning weights to attributes and attribute values based on their occurrence in the query workload.
For example, if $A_1$ occurs 100 times while $A_2$ occurs 50 times, then weight of $A_1=2/3$ and $A_2=1/3$.
We repeat this process for attribute values also.
If an attribute value does not occur in the query workload, we assign a token frequency of 1.
For example, if $A_1=1$ occurs 100 times while $A_2$ occurred none, then their weights are $100/101$ and $1/101$ respectively.
We compute the frequency distribution of the number of predicates from the query workload (such as \# queries with 1, 2, 3, $\ldots$ predicates).
This information is used to perform weighted sampling of the queries by extending the algorithm for the no workload scenario.
This ensures that queries involving popular attributes and attribute values are generated at a higher frequency.
Of course, sampling takes place without replacement so that all the queries in the augmented query workload are distinct.

Next, we assign different weights $w(q_i)$ to the queries $q_i \in Q'$ from the augmented workload to
ensure that the model prioritizes the accuracy of queries from the workload.
\begin{equation}
	w(q_i) =
		\begin{cases}
			1 & \mbox{ if } q_i \in Q \\
			\frac{|Q|}{|Q'|} & \mbox{ if } q_i \not \in Q
		\end{cases}
\end{equation}
We then train the DL model where the penalty for a query $q$ is weighed proportionally to whether it came from the original or the augmented query workload.
\begin{equation}
	\text{q-error}(Q) = \frac{1}{|Q|} \sum_{i=1}^{|Q|}  \mathbf{w(q_i)} \times \left( \frac{s_i}{\widehat{s_i}} + \frac{\widehat{s_i}}{s_i} \right)
\end{equation}

\subsection{Miscellaneous Issues}
\label{subsec:incrementalSupervised}


\stitle{Range Queries.}
The generic nature of our query featurization allows us to transparently include range queries.
Instead of one-hot encoding the attribute values of the point query,
we just need to one-hot encode the specified range.
Note that if an attribute $A_i$ is not specified in the query, we encode the range as the minimum and maximum value from the domain of $A_i$.

\stitle{Incremental Data.}
Our supervised approach does not even look at the data and only uses the query training dataset.
When incremental data arrives, the selectivities of some of these queries would change.
We then train the model on the dataset with the updated selectivites.

\stitle{Incremental Query Workload.}
In this case, it is possible to use the incremental training algorithm as described in Section~\ref{subsec:incrementalMADE}.
We initialize the supervised model with the weights from previous training run instead of random initialization.
We train the model on the new data with a reduced learning rate and a smaller number of epochs.
We also use Dropout regularization technique with probability $p=0.1$ to avoid catastrophic forgetting.

\section{Experiments}
\label{sec:experiments}


\begin{figure*}[ht!]
    \begin{minipage}[t]{0.48\linewidth}
        \centering
        \includegraphics[scale=0.45]{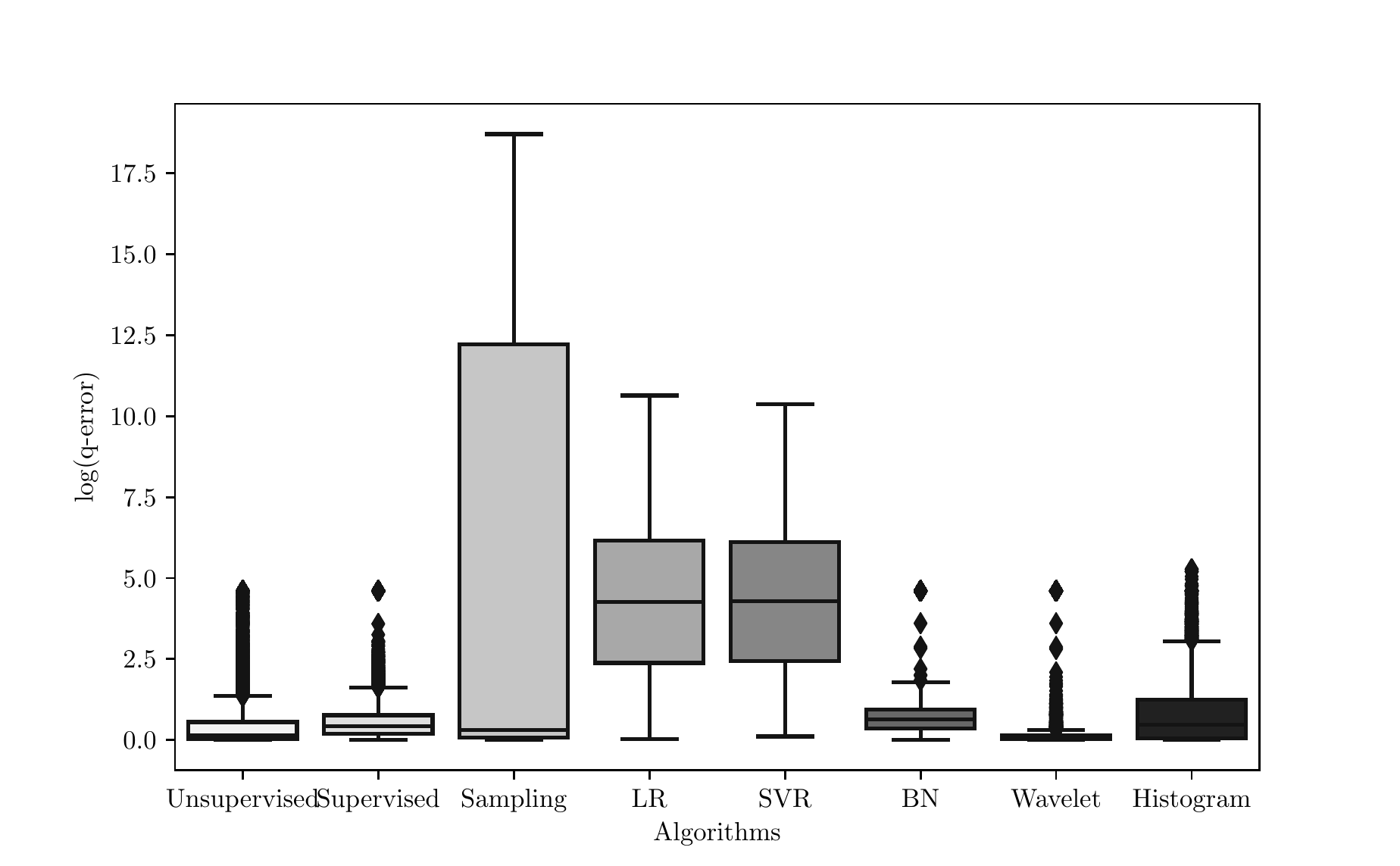}
        \caption{Comparison with Baselines (Census)}
        \label{fig:census_baseline}
    \end{minipage}
    \begin{minipage}[t]{0.48\linewidth}
        \centering
        \includegraphics[scale=0.45]{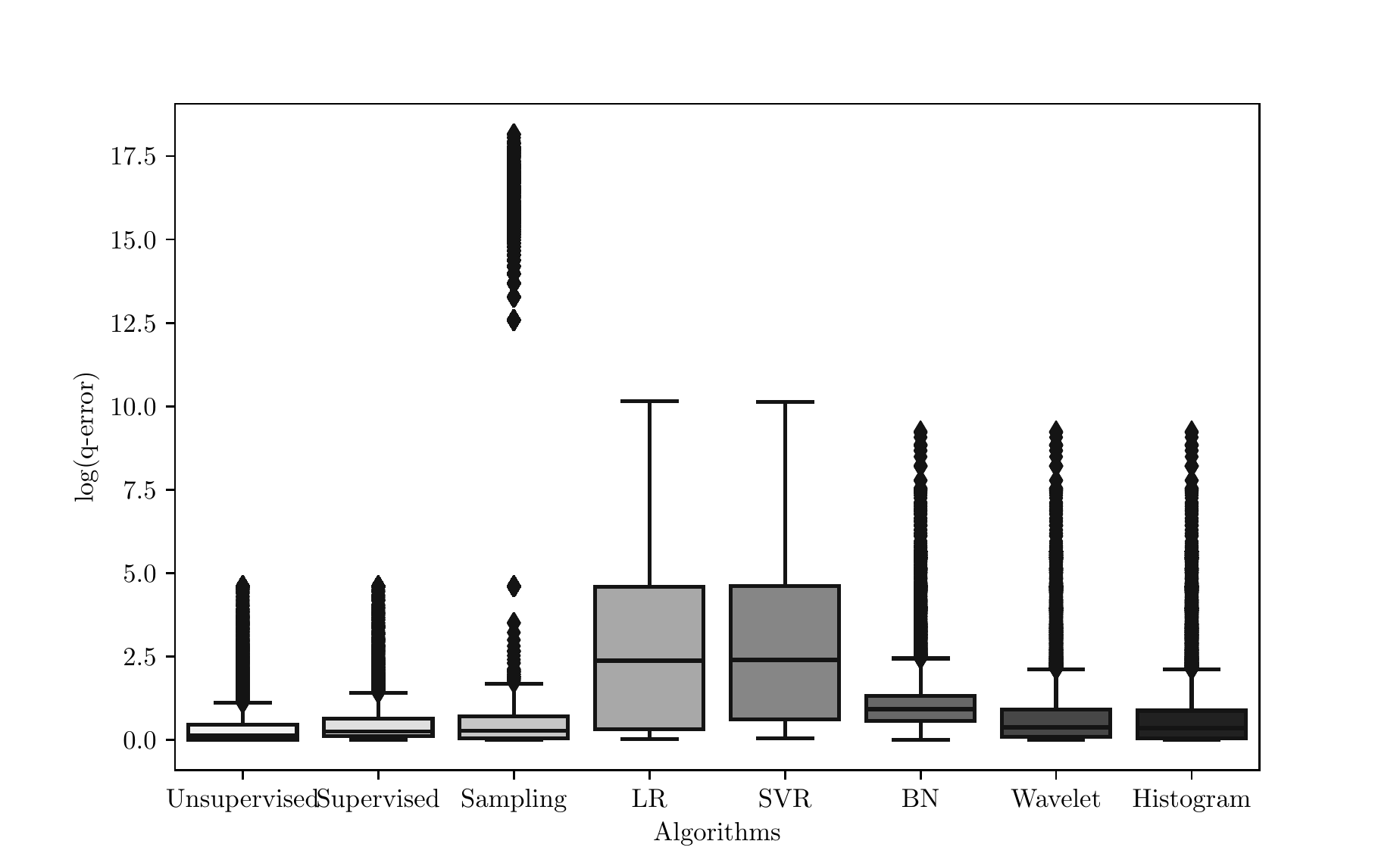}
        \caption{Comparison with Baselines (IMDB)}
        \label{fig:imdb_baseline}
    \end{minipage}
\end{figure*}

\begin{figure*}[ht!]
    \begin{minipage}[t]{0.23\linewidth}
        \centering
        \includegraphics[width =0.95\textwidth]{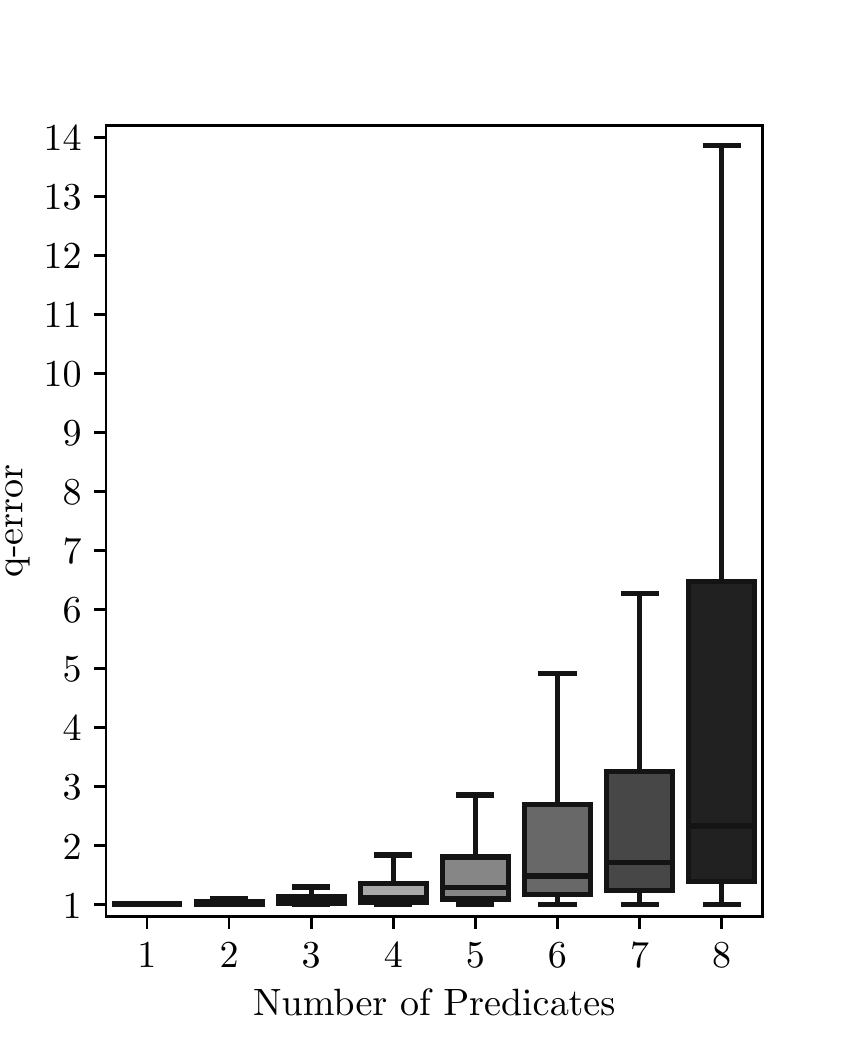}
        \caption{Varying \#Predicates (Unsupervised)}
        \label{fig:census_made_predicates}
    \end{minipage}
    \hspace{1mm}
    \begin{minipage}[t]{0.23\linewidth}
        \centering
        \includegraphics[width =0.95\textwidth]{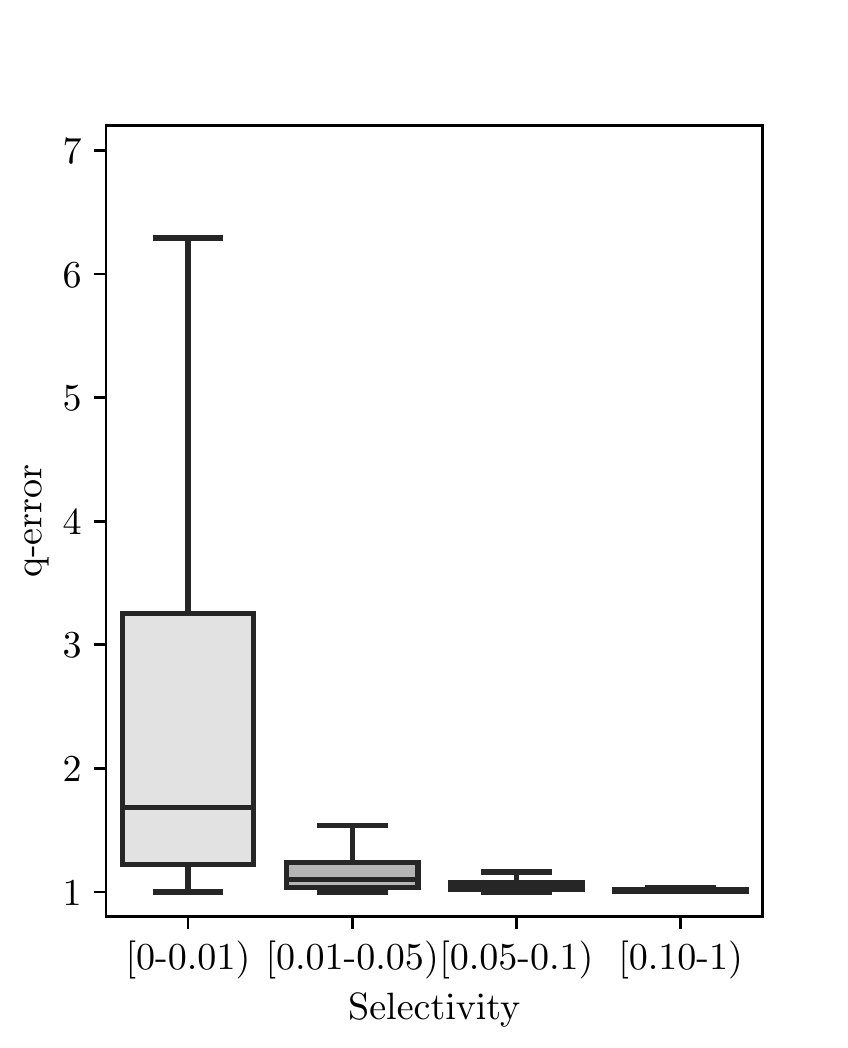}
        \caption{{Varying Selectivity (Unsupervised)}}
        \label{fig:census_made_selectivity}
    \end{minipage}
    \begin{minipage}[t]{0.23\linewidth}
        \centering
        \includegraphics[width =0.95\textwidth]{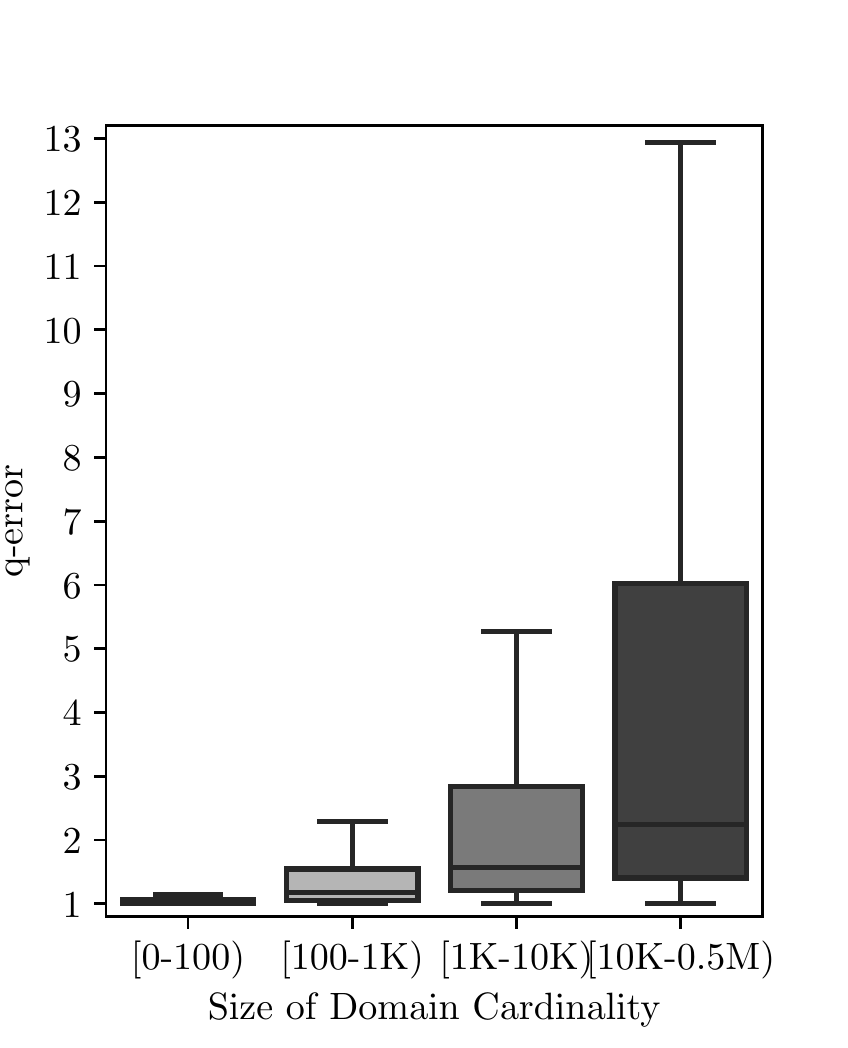}
        \caption{Varying Domain Cardinality (Unsupervised)}
        \label{fig:census_made_cardinality}
    \end{minipage}
    \hspace{1mm}
    \begin{minipage}[t]{0.23\linewidth}
        \centering
        \includegraphics[width =0.95\textwidth]{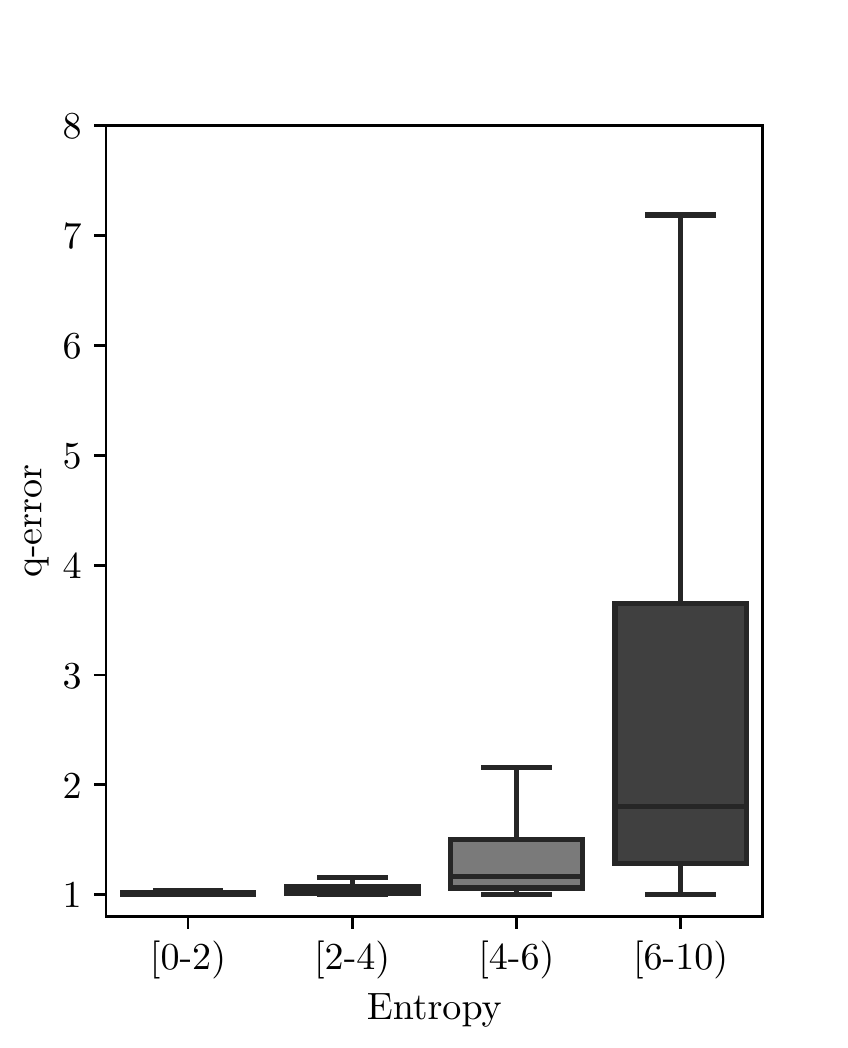}
        \caption{Varying Entropy (Unsupervised)}
        \label{fig:census_made_entropy}
    \end{minipage}
\end{figure*}

\begin{figure*}[!ht]
    \hspace{1mm}
    \begin{minipage}[t]{0.23\linewidth}
        \centering
        \includegraphics[width =0.95\textwidth]{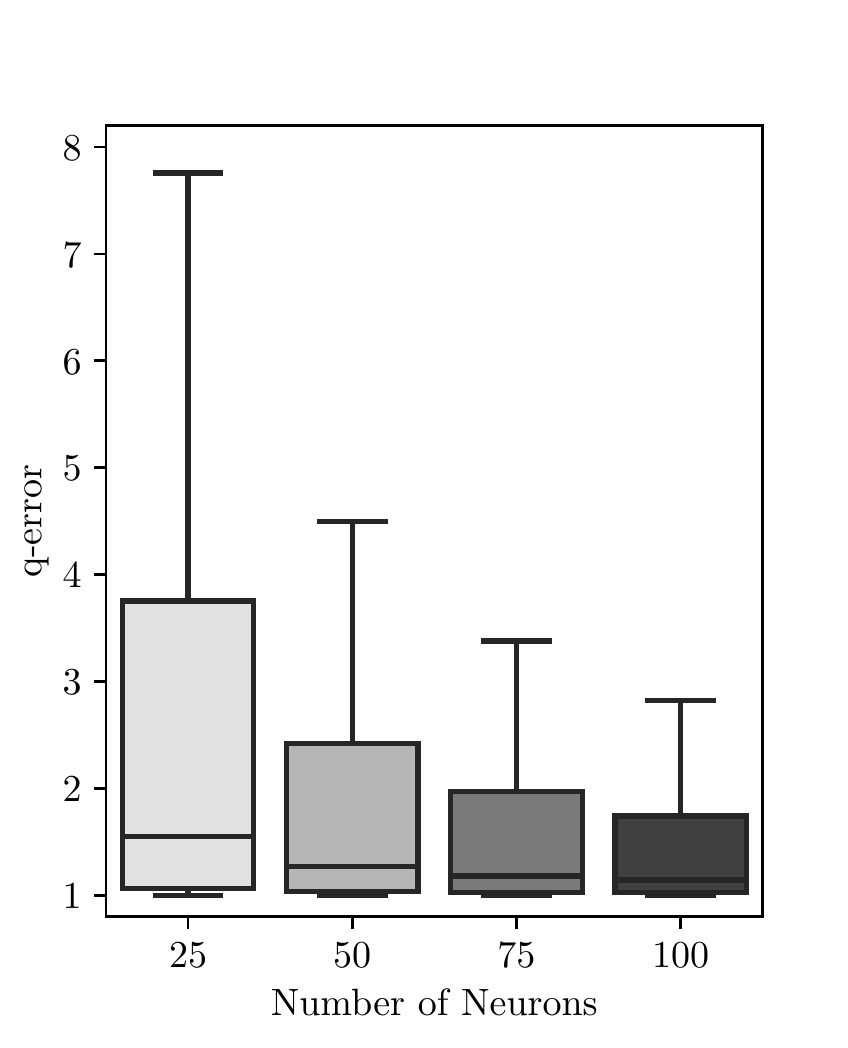}
        \caption{Varying \#Neurons (Unsupervised)}
        \label{fig:census_made_neurons}
    \end{minipage}
    \hspace{1mm}
    \begin{minipage}[t]{0.23\linewidth}
        \centering
        \includegraphics[width =0.95\textwidth]{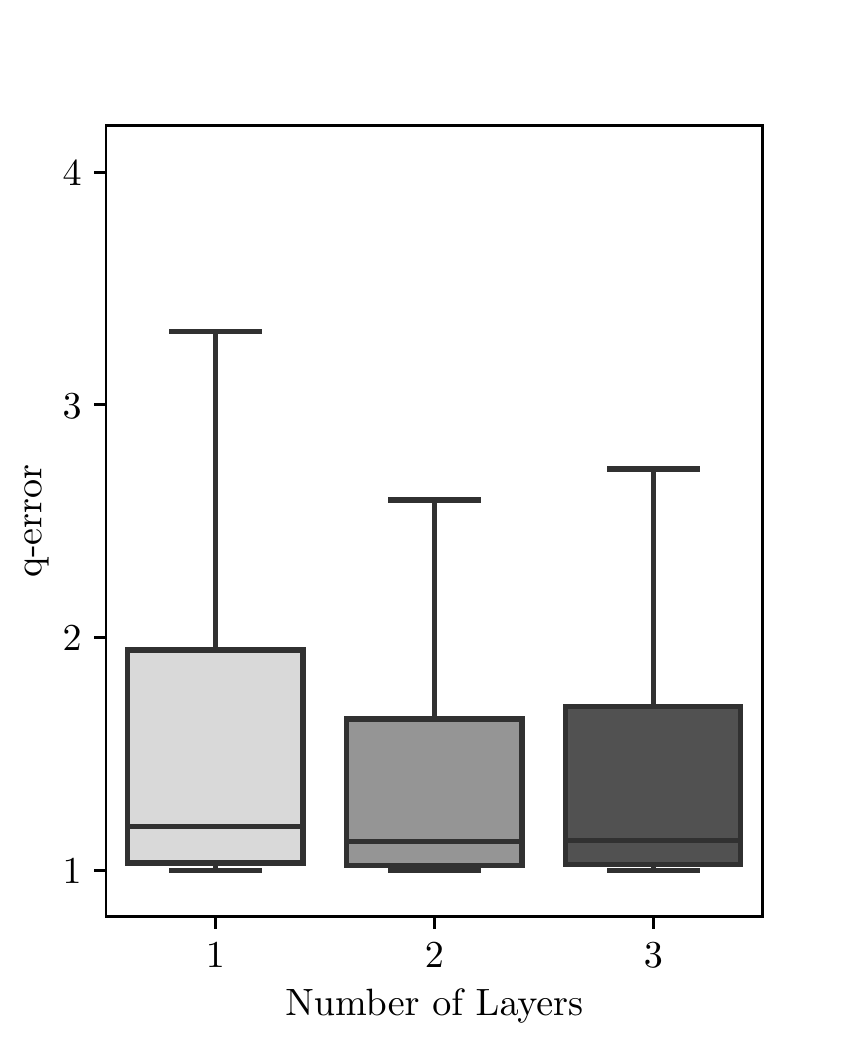}
        \caption{Varying \#Layers (Unsupervised)}
        \label{fig:census_made_layers}
    \end{minipage}
    \hspace{1mm}
    \begin{minipage}[t]{0.23\linewidth}
        \centering
        \includegraphics[width =0.95\textwidth]{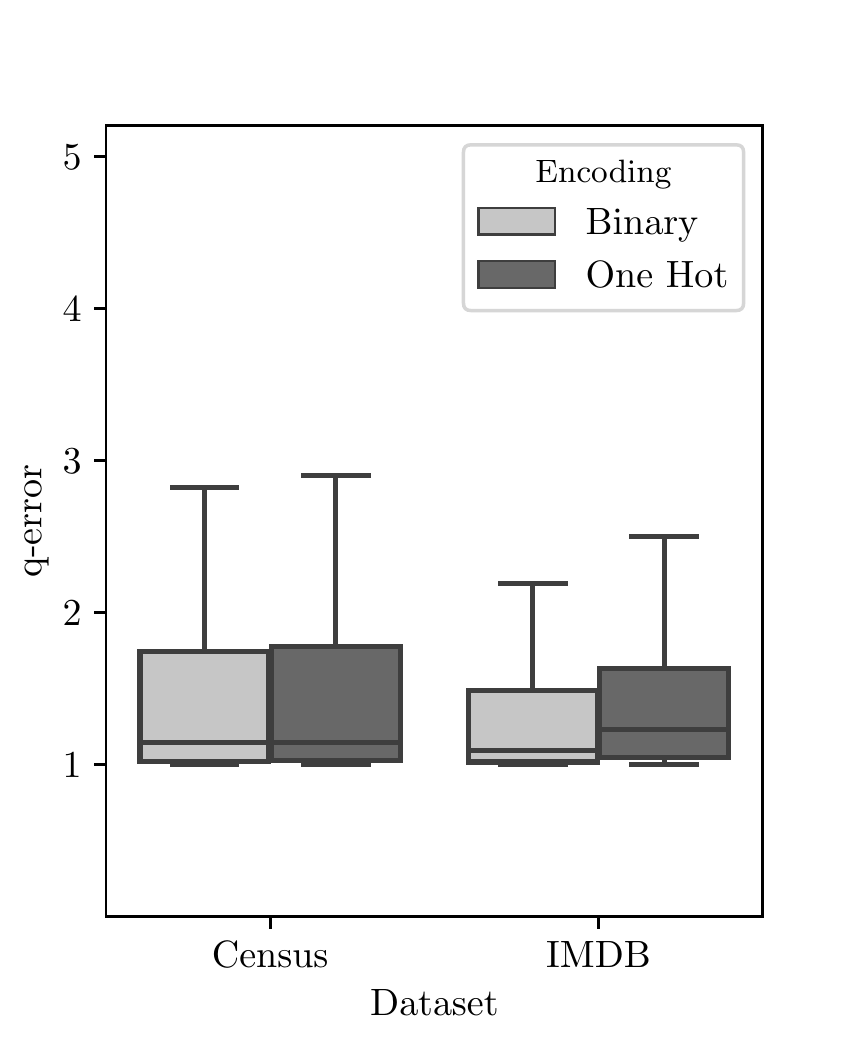}
        \caption{Impact of Encoding (Unsupervised)}
        \label{fig:census_made_encoding}
    \end{minipage}
    \hspace{1mm}
    \begin{minipage}[t]{0.23\linewidth}
        \centering
        \includegraphics[width =0.95\textwidth]{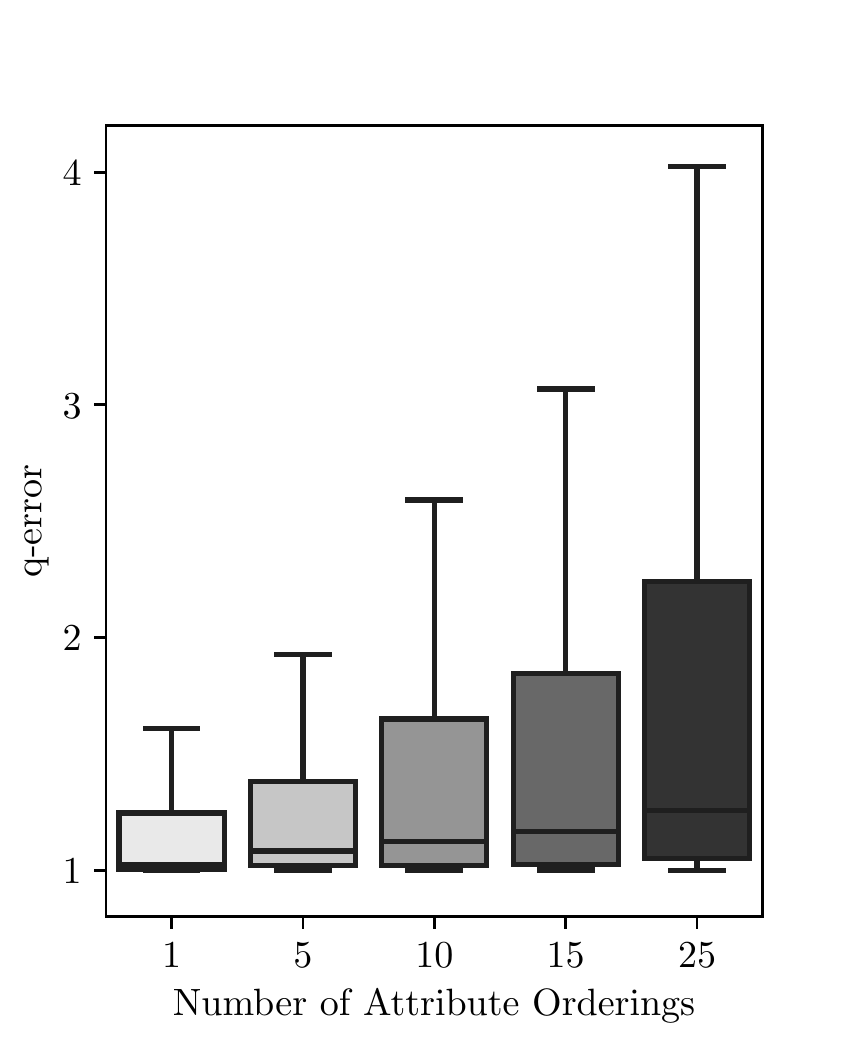}
        \caption{Varying \#Masks (Unsupervised)}
        \label{fig:census_made_masks}
    \end{minipage}
\end{figure*}

In our evaluation, we consider the following key questions:
(1) how do DL based methods compare against prior selectivity estimation approaches commonly used in database systems?
(2) how is the performance of our unsupervised and supervised methods affected by query characteristics
such as number of predicates/attributes, selectivity, size of joint probability distribution and correlated attributes;
(3) how does changing the DL model parameters such as number of neurons, layers and training epochs affects performance?

\subsection{Experimental Setup}

\stitle{Hardware and Platform.}
All our experiments were performed on a NVidia Tesla K80 GPU.
The CPU is a quad-core 2.2 GHz machine with 16 GB of RAM.
We used PyTorch for building the DL models.

\stitle{Datasets.}
We conducted our experiments on two real-world datasets: Census~\cite{getoor2001selectivity} and IMDB from Join Order Benchmark~\cite{leis2015good}.
Both datasets have complex correlated attributes and conditional dependencies.
Selectivity estimation on multiple predicates on these datasets are quite challenging and hence have been extensively used in prior work~\cite{leis2015good}.
The Census dataset has 50K rows, 8 categorical and 6 numerical attributes.
Overall, the IMDB dataset consists of 21 tables with information about movies, actors, directors etc.
For our experiments, we used two large tables Title.akas and Title.basics containing 3.4M and 5.3M tuples with 8 and 9 attributes respectively.

\stitle{Algorithms for Selectivity Estimation.}
The unsupervised model consists of a 2 layer masked autoencoder with 100 neurons in each layer.
Both our algorithms were trained for 100 epochs by default.
We used 1\% sample of IMDB data for training the DL algorithms.
The supervised model consists of 2 fully connected layers with 100 neurons and ReLU activation function.
The final layer has sigmoid activation function to convert the output in the range $[0, 1]$.
The training data consists of 10K queries (see details in Section~\ref{sec:supervised}).

\stitle{Query Workload.}
We compared the algorithms on a test query log of 10K queries.
We generated the log to thoroughly evaluate the performance of the estimators for various facets such as
number of predicates, selectivity, size of joint probability distribution, attribute correlation etc.
Census has 8 categorical attributes thereby creating ${8 \choose 1} + {8 \choose 2} + \ldots {8 \choose 8} = 255$ possible attribute combinations.
The 10K workload was equally allotted such that there are 1250 queries with exactly 1 predicate, 1250 queries with 2 predicates and so on.
There are ${8 \choose i}$ combinations with exactly $i$ attributes.
We allocate 1250 equally between ${8 \choose i}$ combinations.
For a specific attribute combination - say education and marital-status - we pick their values randomly without replacement from their respective domains.

\stitle{Performance Measures.}
We used q-error defined in Section~\ref{sec:preliminaries} for measuring the estimation quality.
Recall that q-error of 1 corresponds to perfect estimate while a q-error of 2 corresponds to an under- or over-estimate by a factor of 2 and so on.
We also use box-plots to concisely describe the results of 10K queries.
The middle line corresponds to the median q-error while the box boundaries correspond to the 25th and 75th percentiles.
The top and bottom whiskers are set to show the 95th and 5th percentiles.

\begin{figure*}[!ht]
    \hspace{1mm}
    \begin{minipage}[t]{0.23\linewidth}
        \centering
        \includegraphics[width =0.95\textwidth]{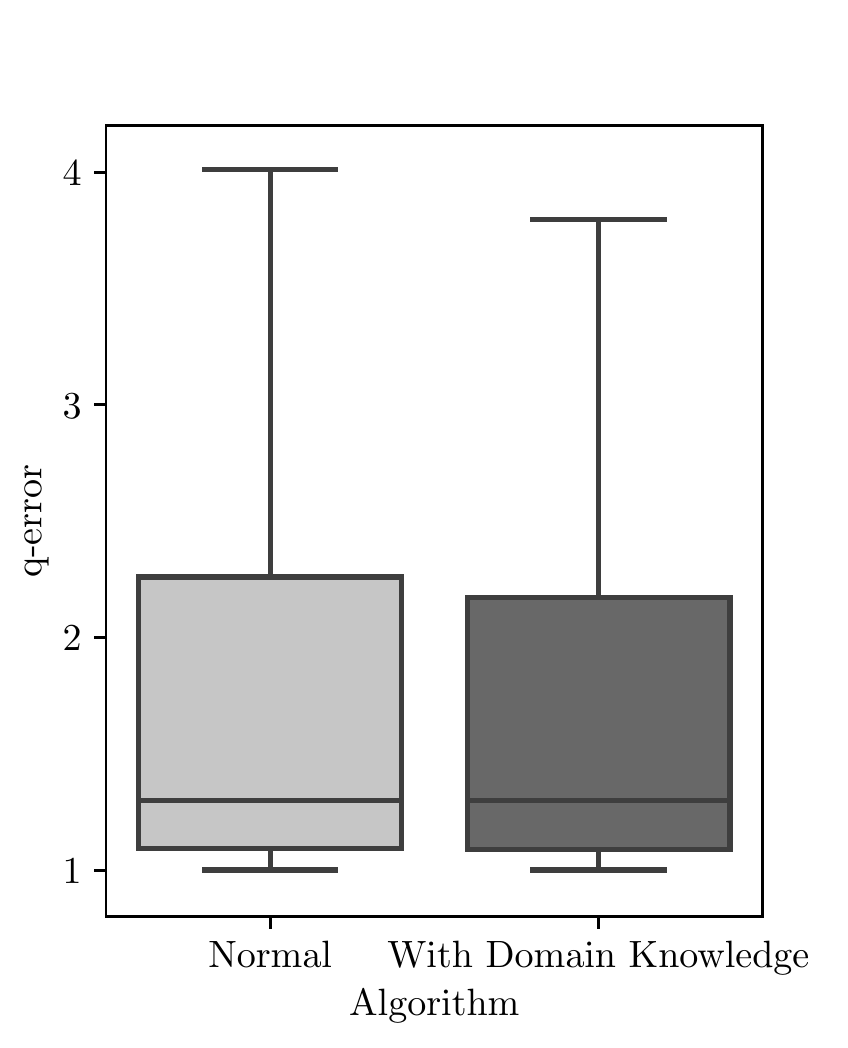}
        \caption{Injecting Domain Knowledge (Unsupervised)}
        \label{fig:census_made_fds}
    \end{minipage}
    \hspace{1mm}
    \begin{minipage}[t]{0.23\linewidth}
        \centering
        \includegraphics[width =0.95\textwidth]{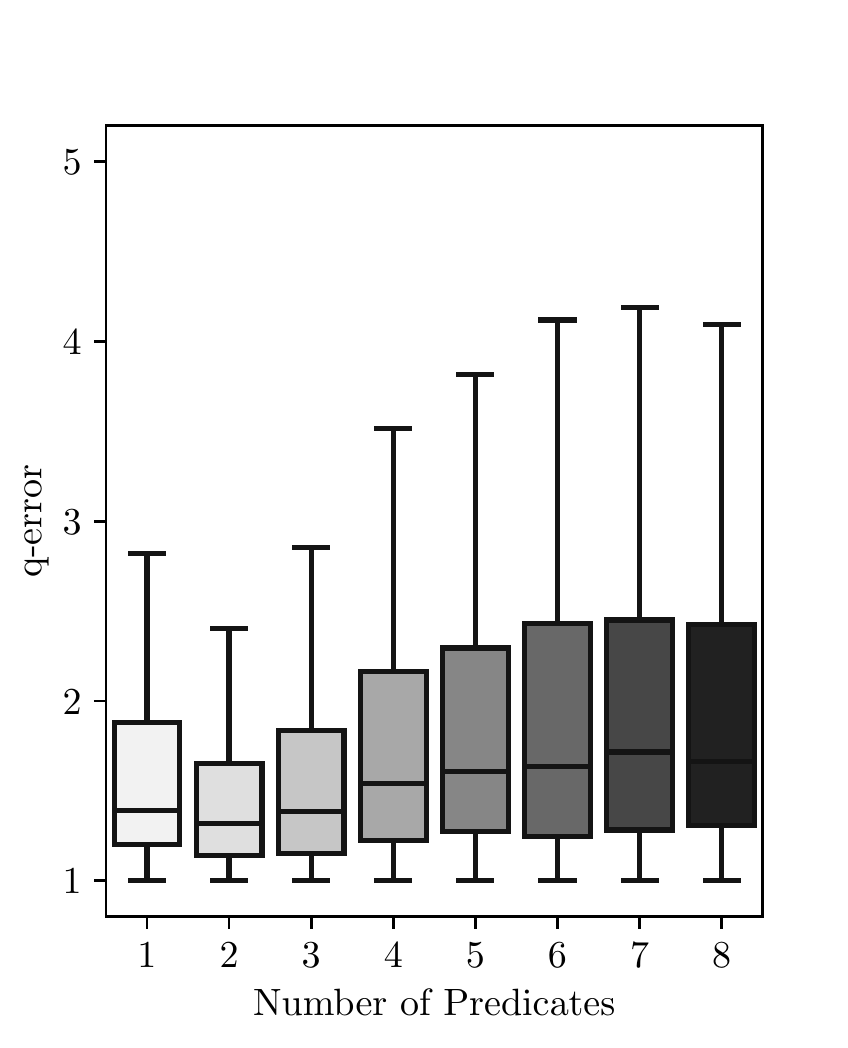}
        \caption{Varying \#Predicates (Supervised)}
        \label{fig:census_supervised_predicates}
    \end{minipage}
    \hspace{1mm}
    \begin{minipage}[t]{0.23\linewidth}
        \centering
        \includegraphics[width =0.95\textwidth]{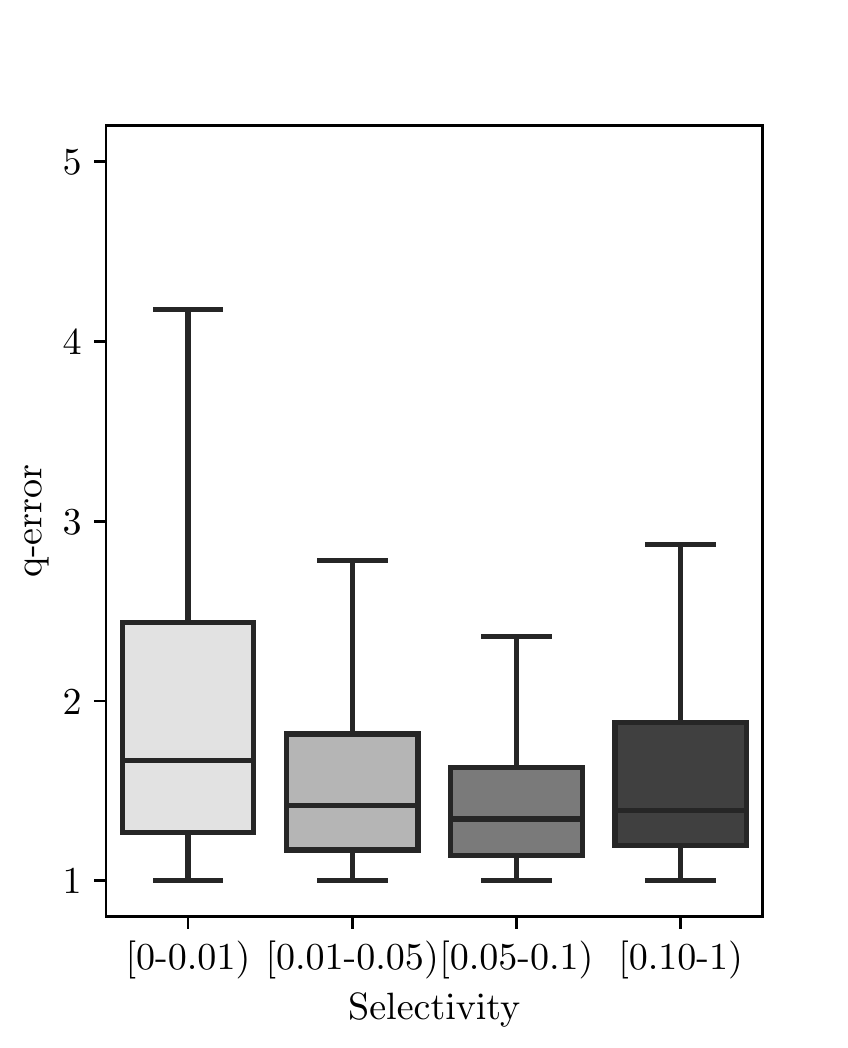}
        \caption{Varying Selectivity (Supervised)}
        \label{fig:census_supervised_selectivity}
    \end{minipage}
    \hspace{1mm}
    \begin{minipage}[t]{0.23\linewidth}
        \centering
        \includegraphics[width =0.95\textwidth]{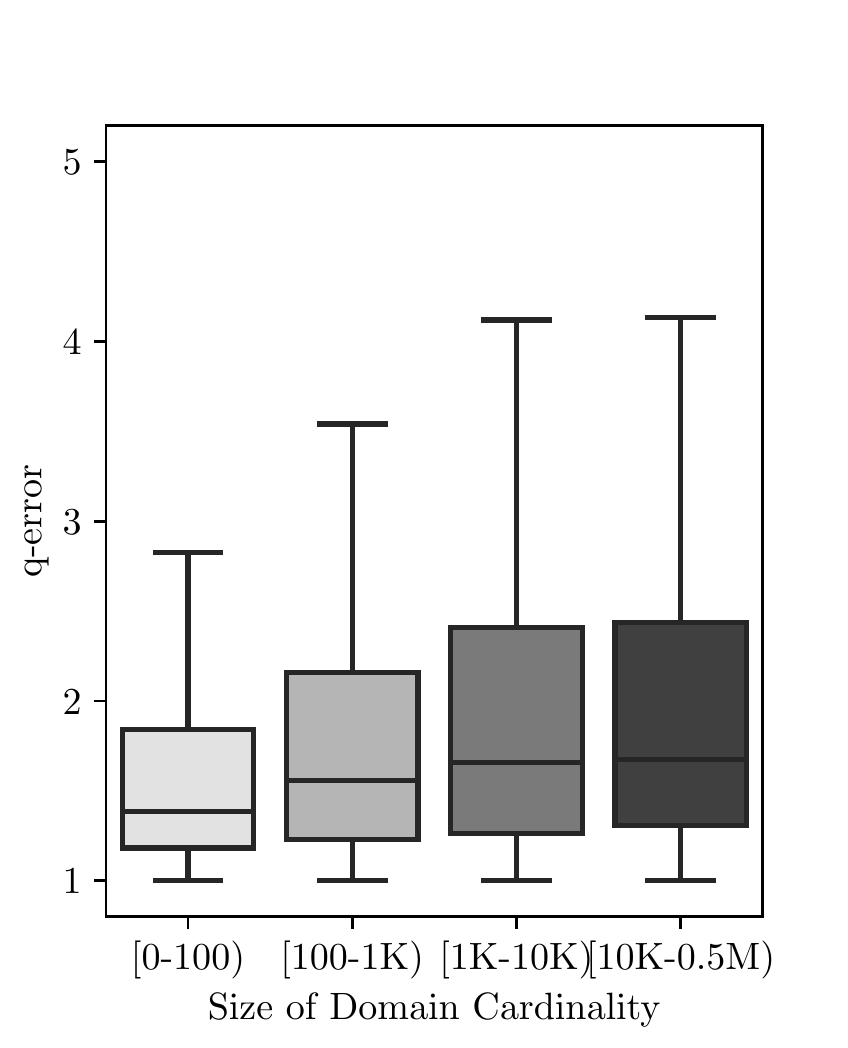}
        \caption{Varying Domain Cardinality (Supervised)}
        \label{fig:census_supervised_cardinality}
    \end{minipage}
\end{figure*}

\begin{figure*}[!ht]
    \hspace{1mm}
    \begin{minipage}[t]{0.23\linewidth}
        \centering
        \includegraphics[width =0.95\textwidth]{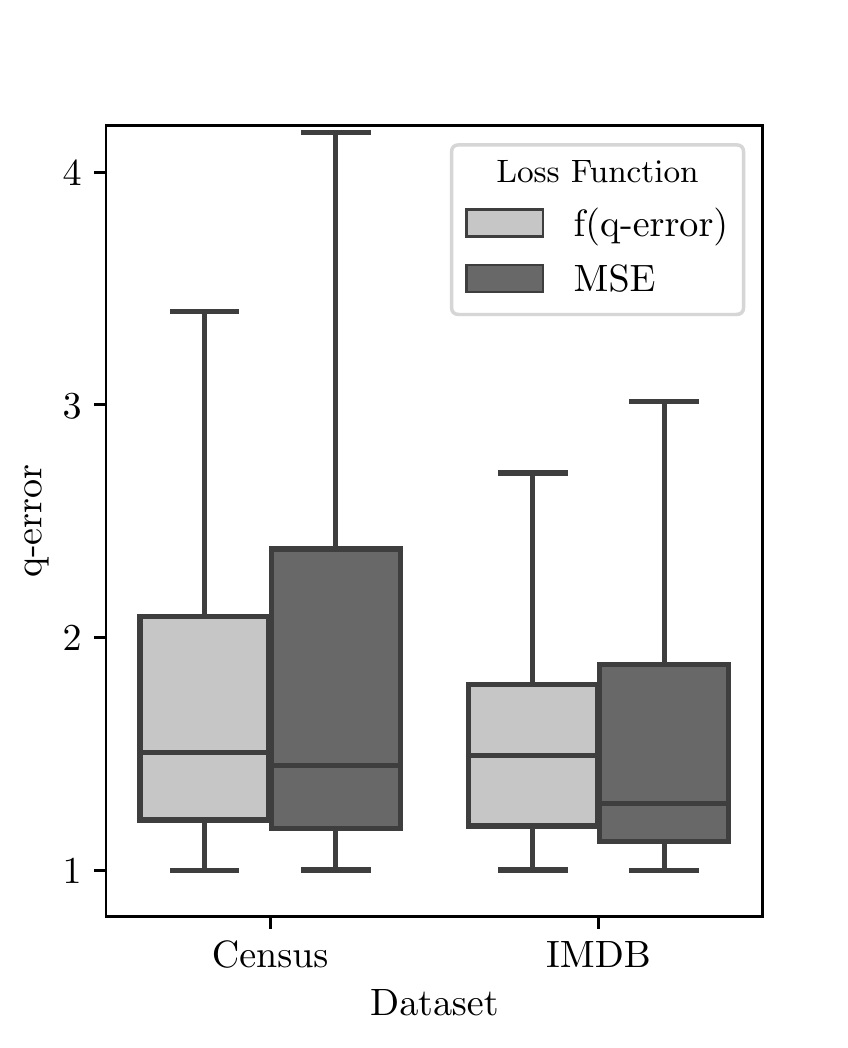}
        \caption{MSE vs Q-Error (Supervised)}
        \label{fig:census_supervised_loss_function}
    \end{minipage}
    \hspace{1mm}
    \begin{minipage}[t]{0.23\linewidth}
        \centering
        \includegraphics[width =0.95\textwidth]{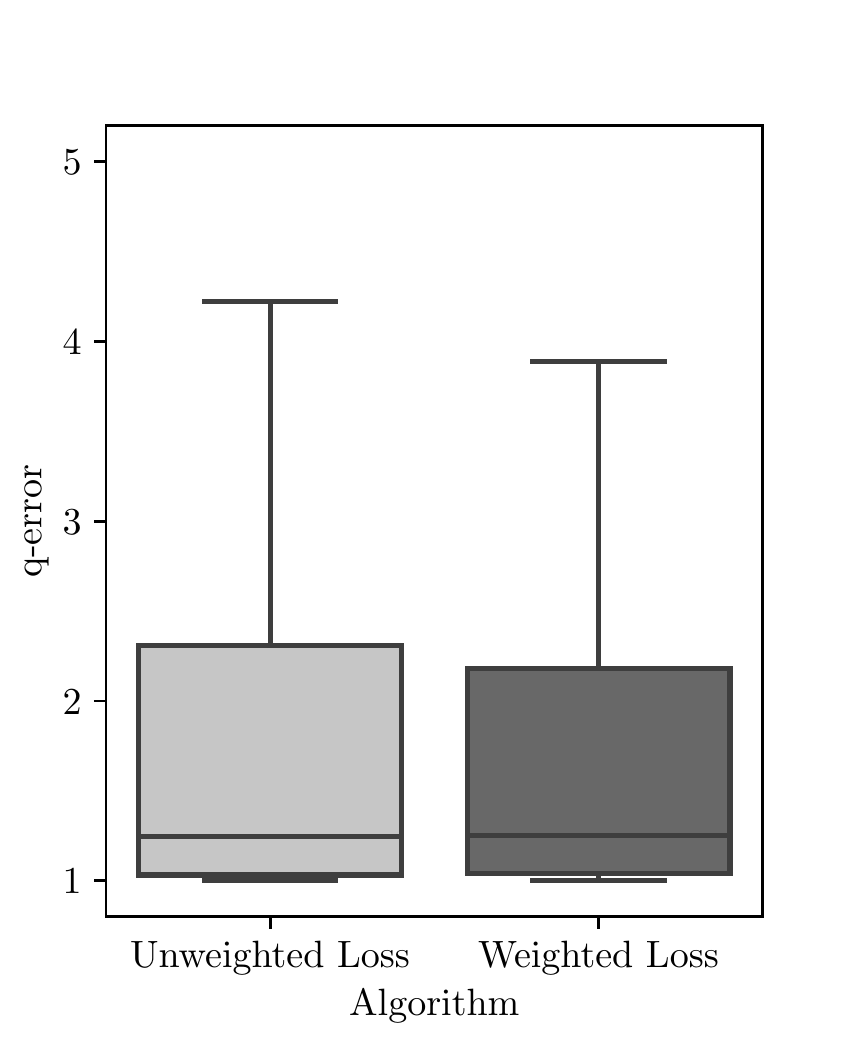}
        \caption{Workload (Unsupervised)}
        \label{fig:census_made_workload}
    \end{minipage}
    \hspace{1mm}
    \begin{minipage}[t]{0.23\linewidth}
        \centering
        \includegraphics[width =0.95\textwidth]{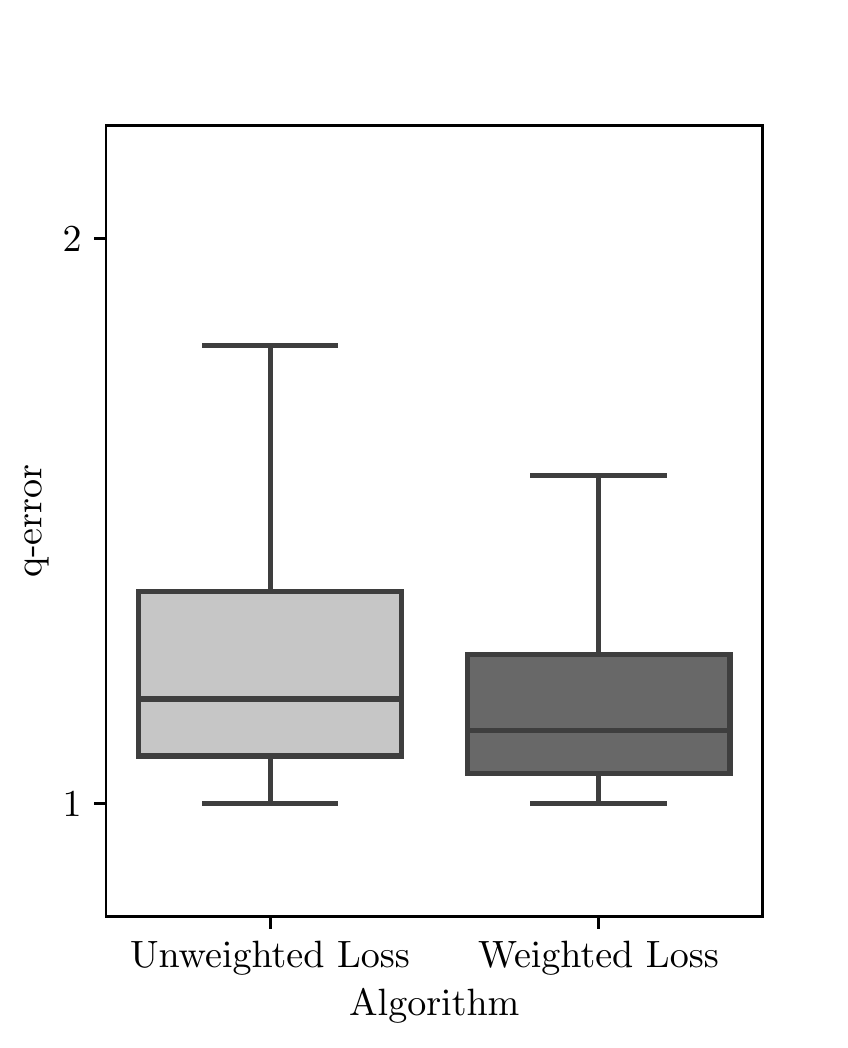}
        \caption{Workload (Supervised)}
        \label{fig:census_supervised_workload}
    \end{minipage}
    \hspace{1mm}
        \begin{minipage}[t]{0.23\linewidth}
    \centering
    \includegraphics[width =1.3\textwidth, height=1.2\textwidth]{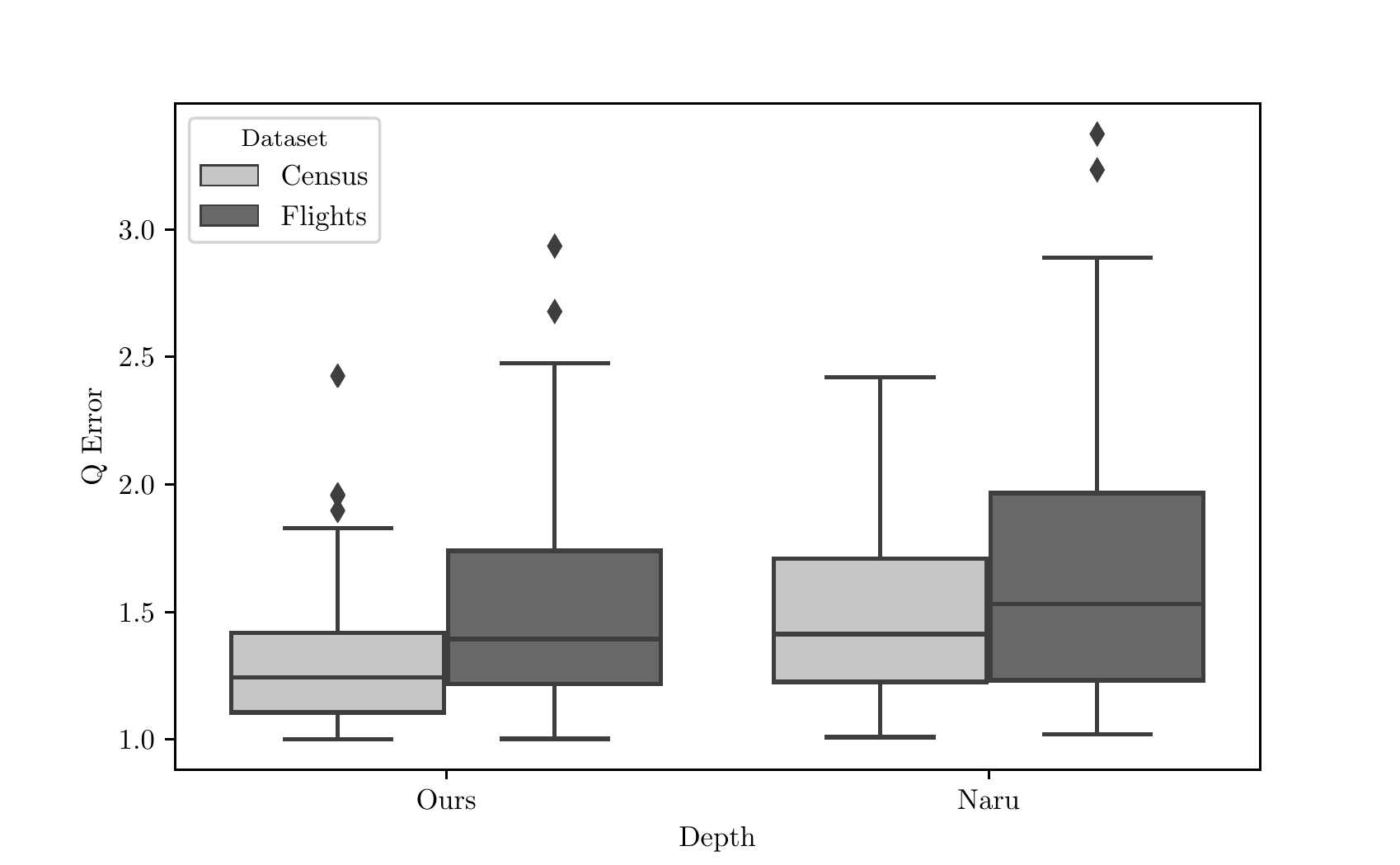}
    \caption{Range Queries}
    \label{fig:range_queries}
    \end{minipage}
\end{figure*}

\subsection{Comparison with Baselines}
\label{subsec:exptResults}

In our first set of experiments, we demonstrate the efficacy of our approaches against popular baseline approaches such as
multi-dimensional histograms~\cite{poosala1996improved, DBLP:conf/sigmod/ThaperGIK02, DBLP:conf/vldb/JagadishKMPSS98, DBLP:conf/vldb/Ioannidis03, DBLP:conf/sigmod/BrunoC04, DBLP:conf/sigmod/BrunoCG01},
wavelets \cite{matias1998wavelet}, Bayesian networks \cite{getoor2001selectivity,tzoumas2011lightweight} and sampling \cite{lipton1990practical}.
Our experiments were conducted on Postgres and leverage the recently introduced multi-column statistics feature from Postgres 10.
We use the TABLESAMPLE command in Postgres 10 with Bernoulli sampling to obtain a 1\% sample.
Haar wavelets are widely used in selectivity estimation and approximate query processing~\cite{matias1998wavelet,cormode2011synopses} as they are accurate and can be computed in linear time.
We used standard Haar wavelet decomposition algorithm described in~\cite{cormode2011synopses} for handling multi-dimensional data.
Finally, we used entropy based discretization~\cite{dougherty1995supervised} for Bayesian networks (denoted as BN) so that it fits into the space budget.
We implemented the algorithm described in~\cite{getoor2001selectivity}.
We also evaluated our approach against Linear Regression (denoted as LR)~\cite{neter1989applied} and Support Vector Regression (denoted as SR)~\cite{smola2004tutorial}
that has been previously used for a related area of query performance monitoring~\cite{akdere2012learning}.
For fair comparison, we ensured that all the selectivity estimators are allocated the same space budget.
For example, our single supervised model for the entire Census dataset requires around 200 KB.
However, we allocate for each multi-dimensional histogram and wavelets for various attribute combinations the same space budget.

Figures~\ref{fig:census_baseline} and~\ref{fig:imdb_baseline} present the results.
The y-axis depicts the q-error in log scale. So for Figures~\ref{fig:census_baseline} and~\ref{fig:imdb_baseline}  a value of 0 corresponds to perfect estimate as $\log 1 = 0$.
We can observe that our DL based approaches dramatically outperform all the prior methods.
The baseline approaches of LR~\cite{neter1989applied} and SVR~\cite{smola2004tutorial} provide inaccurate results; both Census and IMDB exhibit complex correlation and conditional dependencies, which these techniques are unable to adapt to.
The sampling based approach provides good estimates for queries with large/medium selectivities but
dramatically drops off in quality/accuracy for queries with low/very-low selectivities.
Wavelets and histograms provide performance comparable to our methods.
However, this is due to the fact that we disproportionately allocated much more resources for them than our approaches.
Interestingly, the closest baseline is BN that is related in principle to our algorithm.
However, our approaches are superior to BN in both accuracy and time. Specifically our approaches are
2 times more accurate on average for Census and 100 times more accurate for the worst case error. Similar trends hold for IMDB in terms of accuracy.
As a point of reference, it takes one minute to train our approaches on
Census versus 16 minutes for BN (correspondingly 12 minutes of training for our approaches for IMDB versus 516 minutes for BN).
Recall that learning the optimal structure of a BN is very expensive.
In contrast, our approach is much faster due to the use of multiple random attribute orderings.

Our methods begin to dramatically outperform the baseline for queries involving 4 or more predicates.
This is consistent with the expectation that DL approaches are able to learn complex relationships in higher dimensions and depict superior accuracy.
Our results were consistently better across various parameters of interest such as selectivity, number of predicates, size of joint probability table and attribute correlation.
We demonstrate each of them with the following experiments.

\subsection{Unsupervised Density Estimation}
\label{subsec:exptResultsUnsupervised}
We begin by investigating the performance of our masked autoencoder based approach for evaluating selectivity estimation.
There are four key dimensions whose impact must be analyzed.
They include the number of predicates in the query, selectivity of the query, size of joint probability distribution table
and finally the correlation between attributes involved in the query.
Due to space limitations we provide the results for the Census dataset only.
The trends for the IMDB data set were similar to that of Census.

\stitle{Varying \#Predicates in Query.}
Figure~\ref{fig:census_made_predicates} depicts how our unsupervised approach behaves for queries with varying number of predicates.
As expected, the approach is very accurate for queries with small number of predicates.
This is unsurprising as they could be easily learnt by most selectivity estimators.
We can observe however that our estimates are very accurate and within a factor of two even for queries {\em with as much as 7-8 predicates}.
Often queries with large number of predicates have small selectivities and exhibit complex correlations.
Despite those challenges our methods provide very good performance.

\stitle{Varying Query Selectivity.}
Next, we group the queries in the 10K test set based on their selectivity.
Figure~\ref{fig:census_made_selectivity} presents that,
our approach provides very accurate estimates for queries with selectivity of 5\% or more.
Even when the selectivities are low or very low,
our method is still able to provide excellent estimates that are off by a factor of at most 2 for 75\% percent of the query test set.

\stitle{Varying Size of Joint Probability Distribution.}
Recall that the size of the joint probability distribution (JPD) increases exponentially with more and more attributes and/or attributes with large domain cardinality.
Hence, synopses based methods have to make simplifying assumptions such as attribute value independence for compactly representing distributions.
Figure~\ref{fig:census_made_cardinality} demonstrates the performance of our approach.
As expected, our methods can produce very accurate estimates when the size of JPD is small.
Nevertheless, even when the size of JPD is very large (almost 3.8M for Census), it is still off only by a small factor.

\stitle{Varying Attribute Correlation.}
Another major factor is the correlation and dependencies between attributes.
If a set of attributes are correlated, then simplistic approaches such as attribute value independence yield very large estimation errors.
We use entropy to quantify the challenge in concisely modeling the joint distribution.
Intuitively, simple distributions such as uniform have a large entropy while highly correlated attributes have a small entropy.
Figure~\ref{fig:census_made_entropy} shows that our model performs very accurate estimations for small entropy.
This demonstrates that it is able to successfully learn relationships between attribute distributions.

\stitle{Varying Model Hyperparameters.}
Figures~\ref{fig:census_made_neurons} and~\ref{fig:census_made_layers} present the impact of varying the two major hyperparameters.
We begin by varying the number of neurons in each layer from 25 to 100.
We observe that as the number of neurons increases, the q-error decreases with a hint of diminishing returns.
Larger number of neurons increases the model capacity available to learn complex distributions at the cost of increased training time.
Figure~\ref{fig:census_made_layers} demonstrates that increasing the number of layers has a milder impact on performance.
A model with 2 layers is already expressive enough to handle the data distribution of Census.

\stitle{Range Queries.}
We next evaluated the performance of our adaptive importance sampling algorithm for answering range queries with upto 4 range predicates.
We compared our approach with~\cite{yang2019selectivity} that proposed a progressive sampling that also provides unbiased estimates.
We fixed the sampling budget to 500 samples.
Figure~\ref{fig:range_queries} shows that our approach provides slightly better estimates than Naru.

\stitle{Miscellaneous Experiments.}
Figure~\ref{fig:census_made_encoding} presents the impact of the tuple encoding for the two datasets.
For simple datasets such as Census where most of the attributes have small domain cardinality, both approaches provide comparable performance.
However, for datasets such as IMDB where domain cardinalities could be in the range of 10s of thousands, binary encoding outperforms simple one-hot encoding.
Figure~\ref{fig:census_made_masks} shows that
performance drops when the number of random attribute orderings $\kappa$ increases as it also simultaneously increases the possibility of a bad ordering.
Empirically, we found that using a value between $1$ and $5$ provides best results.
Figure~\ref{fig:census_made_fds} presents the impact of injecting domain knowledge appropriately.
For this experiment, we only considered queries whose predicates are a superset of attributes involved in a functional dependency.
We also set the value of $\kappa$ to 1.
The results demonstrate that filtering attribute orderings based on domain knowledge provides a non-negligible improvement.

\subsection{Supervised Density Estimation}
\label{subsec:exptResultsSupervised}

We next evaluate the performance of our supervised selectivity estimator.
Overall, the trends are similar to that of the unsupervised case.
Figure~\ref{fig:census_supervised_predicates} presents the result of varying the number of predicates.
As expected, the q-error increases with increasing number of predicates.
Nevertheless, most of the predicates have a q-error of at most 2.
As discussed in the baseline experiments, our proposed approach outperforms prior selectivity estimation approaches dramatically for queries with large number of predicates.
Figure~\ref{fig:census_supervised_selectivity} demonstrates that our q-error decreases as the query selectivity increases.
Our model has a median q-error of less than 2 even for queries with selectivity less than 1\%.
Figure~\ref{fig:census_supervised_cardinality} shows that as the size of the joint probability distribution increases, the q-error of our model increases; it is still within a factor of 2 however.
A key factor is our proposed algorithm to generate training datasets that provide meaningful and diverse set of queries to train our supervised model.
Figure~\ref{fig:census_supervised_loss_function} presents the impact of the loss function.
As described in Section~\ref{sec:supervised}, directly using q-error as the loss function is desirable to using proxy metrics such as MSE.
For both datasets, the performance of a DL model with q-error is superior to that of MSE.

\section{Related Work}
\label{sec:relWork}

\stitle{Deep Learning for Databases.}
Recently, there has been extensive work on applying techniques from DL for solving challenging database problems.
One of the first work was by Kraska et. al~\cite{kraska2018case} to build fast indices.
The key idea is to use a mixture of neural networks to effectively learn the distribution of data.
This has some obvious connection to our work.
Similar to~\cite{kraska2018case}, we also use a mixture of models for learning the data.
However, the specific nature of the mixture is quite different. 
Specifically, we leverage mixtures to ameliorate the order sensitivity of neural density estimation.
Furthermore, DL autoregressive based approaches often model the data distribution more effectively than those proposed in~\cite{kraska2018case}.
There has been extensive work on using DL techniques including reinforcement learning 
for query optimization (and join order enumeration) 
such as~\cite{ortiz2018learning,marcus2018deep,tzoumas2008reinforcement,krishnan2018learning}.
DL has also been applied to the problem of entity resolution in~\cite{ebraheem2018distributed}.

\stitle{Selectivity Estimation.}
Due to the importance of selectivity estimation, there has been extensive work on accurate estimation.
Popular approaches include sampling~\cite{lipton1990practical}, 
histograms~\cite{poosala1996improved, DBLP:conf/sigmod/ThaperGIK02, DBLP:conf/vldb/JagadishKMPSS98, DBLP:conf/vldb/Ioannidis03, DBLP:conf/sigmod/BrunoC04, DBLP:conf/sigmod/BrunoCG01}, wavelets~\cite{matias1998wavelet}, kernel density estimation~\cite{korn1999range,DBLP:journals/pvldb/KieferHBM17,DBLP:journals/vldb/GunopulosKTD05} and 
graphical models~\cite{getoor2001selectivity,tzoumas2011lightweight}.
Due to its versatility, ML has been explored for the problem of selectivity estimation.
One of the earliest approaches to use neural networks is~\cite{lakshmi1998selectivity}.
While promising, the recently proposed techniques such as neural density estimation are much more accurate.
Another relevant recent work is~\cite{kipf2018learned} that focuses on estimating correlated join selectivities.
It proposes a novel set based DL model but focuses mostly on supervised learning.
In contrast we consider both supervised and unsupervised approaches.

\section{Final Remarks}
\label{sec:futureWork}

In this paper, we investigated the feasibility of applying deep learning based techniques for the fundamental problem of selectivity estimation.
We proposed two complementary approaches that modeled the problem as an supervised and unsupervised learning respectively.
Our extensive experiments showed that the results are very promising and can address some of the pain points of popular selectivity estimators.
There are a number of promising avenues  to explore.
For one, how to extend the selectivity estimators over single tables to multiple tables involving correlated joins.
Another intriguing direction is to investigate the possibility of other deep generative models such as 
deep belief networks (DBN), variational auto encoders (VAE) and generative adversarial networks (GANs) for the purpose of selectivity estimation.

\balance
\interlinepenalty=10000
\bibliographystyle{abbrv}
\bibliography{SelEst}

\begin{thebibliography}{10}

\bibitem{akdere2012learning}
M.~Akdere, U.~{\c{C}}etintemel, M.~Riondato, E.~Upfal, and S.~B. Zdonik.
\newblock Learning-based query performance modeling and prediction.
\newblock In {\em ICDE}, pages 390--401. IEEE, 2012.

\bibitem{DBLP:conf/sigmod/BrunoC04}
N.~Bruno and S.~Chaudhuri.
\newblock Conditional selectivity for statistics on query expressions.
\newblock In {\em SIGMOD}, pages 311--322, 2004.

\bibitem{DBLP:conf/sigmod/BrunoCG01}
N.~Bruno, S.~Chaudhuri, and L.~Gravano.
\newblock Stholes: {A} multidimensional workload-aware histogram.
\newblock In {\em SIGMOD}, pages 211--222, 2001.

\bibitem{changyong2014log}
F.~Changyong, W.~Hongyue, L.~Naiji, C.~Tian, H.~Hua, L.~Ying, et~al.
\newblock Log-transformation and its implications for data analysis.
\newblock {\em Shanghai archives of psychiatry}, 26(2):105, 2014.

\bibitem{cormode2011synopses}
G.~Cormode, M.~Garofalakis, P.~J. Haas, C.~Jermaine, et~al.
\newblock Synopses for massive data: Samples, histograms, wavelets, sketches.
\newblock {\em Foundations and Trends in Databases}, 4(1--3):1--294, 2011.

\bibitem{dougherty1995supervised}
J.~Dougherty, R.~Kohavi, and M.~Sahami.
\newblock Supervised and unsupervised discretization of continuous features.
\newblock In {\em Machine Learning Proceedings 1995}, pages 194--202. Elsevier,
  1995.

\bibitem{rangeQueriesML}
A.~Dutt, C.~Wang, A.~Nazi, S.~Kandula, V.~Narasayya, and S.~Chaudhuri.
\newblock Selectivity estimation for range predicates using lightweight models.
\newblock {\em {PVLDB}}, 12(9):1044 -- 1057, 2019.

\bibitem{ebraheem2018distributed}
M.~Ebraheem, S.~Thirumuruganathan, S.~Joty, M.~Ouzzani, and N.~Tang.
\newblock Distributed representations of tuples for entity resolution.
\newblock {\em PVLDB}, 11(11):1454--1467, 2018.

\bibitem{germain2015made}
M.~Germain, K.~Gregor, I.~Murray, and H.~Larochelle.
\newblock Made: Masked autoencoder for distribution estimation.
\newblock In {\em ICML}, pages 881--889, 2015.

\bibitem{getoor2001selectivity}
L.~Getoor, B.~Taskar, and D.~Koller.
\newblock Selectivity estimation using probabilistic models.
\newblock In {\em ACM SIGMOD Record}, volume~30, 2001.

\bibitem{Goodfellow-et-al-2016}
I.~Goodfellow, Y.~Bengio, and A.~Courville.
\newblock {\em Deep Learning}.
\newblock MIT Press, 2016.
\newblock \url{http://www.deeplearningbook.org}.

\bibitem{goodfellow2013empirical}
I.~J. Goodfellow, M.~Mirza, D.~Xiao, A.~Courville, and Y.~Bengio.
\newblock An empirical investigation of catastrophic forgetting in
  gradient-based neural networks.
\newblock {\em arXiv preprint arXiv:1312.6211}, 2013.

\bibitem{DBLP:conf/icml/GregorDMBW14}
K.~Gregor, I.~Danihelka, A.~Mnih, C.~Blundell, and D.~Wierstra.
\newblock Deep autoregressive networks.
\newblock In {\em {ICML}}, pages 1242--1250, 2014.

\bibitem{DBLP:journals/vldb/GunopulosKTD05}
D.~Gunopulos, G.~Kollios, V.~J. Tsotras, and C.~Domeniconi.
\newblock Selectivity estimators for multidimensional range queries over real
  attributes.
\newblock {\em {VLDB} J.}, 14(2):137--154, 2005.

\bibitem{DBLP:conf/vldb/Ioannidis03}
Y.~E. Ioannidis.
\newblock The history of histograms (abridged).
\newblock In {\em {VLDB}}, 2003.

\bibitem{DBLP:conf/vldb/JagadishKMPSS98}
H.~V. Jagadish, N.~Koudas, S.~Muthukrishnan, V.~Poosala, K.~C. Sevcik, and
  T.~Suel.
\newblock Optimal histograms with quality guarantees.
\newblock In {\em VLDB}, pages 275--286, 1998.

\bibitem{DBLP:journals/pvldb/KieferHBM17}
M.~Kiefer, M.~Heimel, S.~Bre{\ss}, and V.~Markl.
\newblock Estimating join selectivities using bandwidth-optimized kernel
  density models.
\newblock {\em {PVLDB}}, 10(13):2085--2096, 2017.

\bibitem{Kingma2015AdamAM}
D.~P. Kingma and J.~Ba.
\newblock Adam: A method for stochastic optimization.
\newblock {\em ICLR}, abs/1412.6980, 2015.

\bibitem{kipf2018learned}
A.~Kipf, T.~Kipf, B.~Radke, V.~Leis, P.~Boncz, and A.~Kemper.
\newblock Learned cardinalities: Estimating correlated joins with deep
  learning.
\newblock {\em arXiv preprint arXiv:1809.00677}, 2018.

\bibitem{korn1999range}
F.~Korn, T.~Johnson, and H.~Jagadish.
\newblock Range selectivity estimation for continuous attributes.
\newblock In {\em ssdbm}, page 244. IEEE, 1999.

\bibitem{kraska2018case}
T.~Kraska, A.~Beutel, E.~H. Chi, J.~Dean, and N.~Polyzotis.
\newblock The case for learned index structures.
\newblock In {\em SIGMOD}, pages 489--504. ACM, 2018.

\bibitem{krishnan2018learning}
S.~Krishnan, Z.~Yang, K.~Goldberg, J.~Hellerstein, and I.~Stoica.
\newblock Learning to optimize join queries with deep reinforcement learning.
\newblock {\em arXiv preprint arXiv:1808.03196}, 2018.

\bibitem{lakshmi1998selectivity}
S.~Lakshmi and S.~Zhou.
\newblock Selectivity estimation in extensible databases-a neural network
  approach.
\newblock In {\em VLDB}, pages 623--627, 1998.

\bibitem{leis2015good}
V.~Leis, A.~Gubichev, A.~Mirchev, P.~Boncz, A.~Kemper, and T.~Neumann.
\newblock How good are query optimizers, really?
\newblock {\em PVLDB}, 9(3), 2015.

\bibitem{leis2017cardinality}
V.~Leis, B.~Radke, A.~Gubichev, A.~Kemper, and T.~Neumann.
\newblock Cardinality estimation done right: Index-based join sampling.
\newblock In {\em CIDR}, 2017.

\bibitem{li2018learning}
Z.~Li and D.~Hoiem.
\newblock Learning without forgetting.
\newblock {\em IEEE TPAMI}, 40(12):2935--2947, 2018.

\bibitem{lipton1990practical}
R.~J. Lipton, J.~F. Naughton, and D.~A. Schneider.
\newblock {\em Practical selectivity estimation through adaptive sampling},
  volume~19.
\newblock ACM, 1990.

\bibitem{marcus2018deep}
R.~Marcus and O.~Papaemmanouil.
\newblock Deep reinforcement learning for join order enumeration.
\newblock {\em arXiv preprint arXiv:1803.00055}, 2018.

\bibitem{matias1998wavelet}
Y.~Matias, J.~S. Vitter, and M.~Wang.
\newblock Wavelet-based histograms for selectivity estimation.
\newblock In {\em ACM SIGMOD Record}, volume~27, 1998.

\bibitem{muller2018improved}
M.~M{\"u}ller, G.~Moerkotte, and O.~Kolb.
\newblock Improved selectivity estimation by combining knowledge from sampling
  and synopses.
\newblock {\em PVLDB}, 11(9):1016--1028, 2018.

\bibitem{murphy2012machine}
K.~P. Murphy.
\newblock {\em Machine learning: a probabilistic perspective}.
\newblock MIT press, 2012.

\bibitem{neter1989applied}
J.~Neter, W.~Wasserman, and M.~H. Kutner.
\newblock Applied linear regression models.
\newblock 1989.

\bibitem{ortiz2018learning}
J.~Ortiz, M.~Balazinska, J.~Gehrke, and S.~S. Keerthi.
\newblock Learning state representations for query optimization with deep
  reinforcement learning.
\newblock {\em arXiv preprint arXiv:1803.08604}, 2018.

\bibitem{poosala1996improved}
V.~Poosala, P.~J. Haas, Y.~E. Ioannidis, and E.~J. Shekita.
\newblock Improved histograms for selectivity estimation of range predicates.
\newblock In {\em ACM Sigmod Record}, volume~25, 1996.

\bibitem{poosala1997selectivity}
V.~Poosala and Y.~E. Ioannidis.
\newblock Selectivity estimation without the attribute value independence
  assumption.
\newblock In {\em VLDB}, volume~97, pages 486--495, 1997.

\bibitem{DBLP:conf/icml/RifaiVMGB11}
S.~Rifai, P.~Vincent, X.~Muller, X.~Glorot, and Y.~Bengio.
\newblock Contractive auto-encoders: Explicit invariance during feature
  extraction.
\newblock In {\em {ICML}}, pages 833--840, 2011.

\bibitem{smola2004tutorial}
A.~J. Smola and B.~Sch{\"o}lkopf.
\newblock A tutorial on support vector regression.
\newblock {\em Statistics and computing}, 14(3):199--222, 2004.

\bibitem{srivastava2014dropout}
N.~Srivastava, G.~Hinton, A.~Krizhevsky, I.~Sutskever, and R.~Salakhutdinov.
\newblock Dropout: a simple way to prevent neural networks from overfitting.
\newblock {\em JMLR}, 15(1):1929--1958, 2014.

\bibitem{DBLP:conf/sigmod/ThaperGIK02}
N.~Thaper, S.~Guha, P.~Indyk, and N.~Koudas.
\newblock Dynamic multidimensional histograms.
\newblock In {\em SIGMOD}, pages 428--439, 2002.

\bibitem{tzoumas2011lightweight}
K.~Tzoumas, A.~Deshpande, and C.~S. Jensen.
\newblock Lightweight graphical models for selectivity estimation without
  independence assumptions.
\newblock {\em PVLDB}, 4(11):852--863, 2011.

\bibitem{tzoumas2008reinforcement}
K.~Tzoumas, T.~Sellis, and C.~S. Jensen.
\newblock A reinforcement learning approach for adaptive query processing.
\newblock {\em History}, 2008.

\bibitem{DBLP:journals/corr/UriaML13}
B.~Uria, I.~Murray, and H.~Larochelle.
\newblock {NADE:} the real-valued neural autoregressive density-estimator.
\newblock {\em CoRR}, abs/1306.0186, 2013.

\bibitem{DBLP:conf/icml/VincentLBM08}
P.~Vincent, H.~Larochelle, Y.~Bengio, and P.~Manzagol.
\newblock Extracting and composing robust features with denoising autoencoders.
\newblock In {\em {ICML}}, 2008.

\bibitem{yang2019selectivity}
Z.~Yang, E.~Liang, A.~Kamsetty, C.~Wu, Y.~Duan, X.~Chen, P.~Abbeel, J.~M.
  Hellerstein, S.~Krishnan, and I.~Stoica.
\newblock Selectivity estimation with deep likelihood models.
\newblock {\em arXiv preprint arXiv:1905.04278}, 2019.

\end{thebibliography}

\end{document}